\documentclass[fleqn,10pt]{wlscirep}
\usepackage[utf8]{inputenc}
\usepackage[T1]{fontenc}
\usepackage{lineno}
\usepackage{array}

\usepackage{verbatim}
\usepackage{amsmath}
\usepackage{amssymb}
\usepackage{wasysym}
\usepackage{graphicx}
\usepackage{tabularx}
\usepackage{xcolor}
\usepackage{bm}          
\usepackage{pifont} 
\usepackage{dcolumn}
\usepackage{float}
\usepackage{subfigure,textcomp,epsfig,siunitx}
\usepackage{array}
\usepackage{longtable}
\usepackage{bm}
\usepackage{booktabs}

\usepackage{caption}
\usepackage{subfigure,textcomp,epsfig,siunitx}
\usepackage{array}
\usepackage{longtable}
\usepackage{bm}
\usepackage{booktabs}

\title{HEX: a complete Database of High-Pressure Elemental Crystal Structures}

\author[1,2,3,*]{Federico Giannessi}
\author[3,4,5]{Simone Di Cataldo}
\author[5,6,7]{Santanu Saha}
\author[2,3]{Lilia Boeri}
\affil[1]{Dipartimento di Scienze Fisiche e Chimiche, Universit\`a degli Studi dell'Aquila, Via Vetoio 40, 67100 L'Aquila, Italy}
\affil[2]{Enrico Fermi Research Center, Via Panisperna 89 A, 00184, Rome, Italy}
\affil[3]{Dipartimento di Fisica, Sapienza Universit\`a di Roma, 00185 Rome, Italy}
\affil[4]{Institut f\"{u}r Festk\"{o}rperphysik, Wien University of Technology, 1040 Wien, Austria}
\affil[5]{Institute of Theoretical and Computational Physics, Graz University of Technology, NAWI Graz, 8010 Graz, Austria}
\affil[6]{Department of Physics, University of Oxford, Parks Rd, Oxford OX1 3PU, United Kingdom}
\affil[7]{Institut de Recherche sur les Céramiques (IRCER), UMR CNRS 7315-Université de Limoges, Limoges 87068, France}

\affil[*]{corresponding author(s): Federico Giannessi (federico.giannessi@graduate.univaq.it)}

\begin{abstract}

This paper introduces the {\em HEX (High-pressure Elemental Xstals)} database, a complete database of the ground-state crystal structures of the first 57 elements of the periodic table, from H to La, at 0, 100, 200 and 300 GPa.
HEX aims to provide a unified reference for high-pressure research,
by compiling all available experimental information on elements at high pressure, and complementing it with the results of
accurate evolutionary crystal structure prediction runs based on Density Functional Theory. 
Besides offering a much-needed reference, our work also serves as a benchmark of the accuracy of current \textit{ab-initio} methods for crystal structure prediction. We find that, in 98 $\%$ of the cases in which experimental information is available, \textit{ab-initio} crystal structure prediction yields structures
which either coincide or are degenerate in enthalpy to within 300 K
 with experimental ones.
The main manuscript contains synthetic tables and figures, while the Crystallographic Information File (cif) for all structures can be downloaded from \textcolor{red}{figshare online repository}.

\end{abstract}
\begin{document}

\flushbottom
\maketitle

\section*{Background \& Summary}

The advent of  21st century marks a pivotal moment for high-pressure research:  advancements diamond anvil cells
design and \textit{in-situ} characterization techniques\cite{Jayaraman1983,FLORESLIVAS20201,Weir1965-xa,Shen2017-ts}
gave access to the realm of multi-megabar pressures, revealing
unexpected and fascinating phenomena,
such as high-temperature conventional superconductivity in H$_3$S\cite{Drozdov2015,Duan_SciRep_2014_SH}, LaH$_{10}$\cite{Eremets_Nature_2019_LaH,Hemley_PRL_2019_LaH} and other superhydrides,~\cite{FLORESLIVAS20201} metal-insulator transition in elemental sodium\cite{Ma2009-ey}, self-ionization of boron\cite{Oganov2009-uv}, electride
behavior in alkali metals\cite{Lithium_Hanfland_2000}, noble-gas solids\cite{Dong2017-kl}, etc.. 

Until the turn of the century,  knowledge on the
behaviour of matter at high pressure was limited and based on indirect evidence. The general expectation was that all matter would
tend to become homogeneous and metallic  to maximize the electronic kinetic energy.
However, experiments over the last 30 years revealed a much more varied behaviour defying this na\"{i}ve expectation. Compounds at high pressures often 
adopt exotic crystal structures, whose stoichiometries, motifs and moieties defy fundamental chemical concepts, such as valence and electronegativity,
which govern the behaviour of matter at ambient pressure.~\cite{Pauling1932-qb} The most striking examples of this so-called \emph{forbidden chemistry} are highlighted in 
several excellent review papers,~\cite{mcmahon_RSC_2006,Hoffmann_chemical_imagination_2007,Zhang_materialHP_NRM_2017,Mao_RMP_HP_2018,miao_chemistry_HP_review} which also offer a glimpse on the underlying physical mechanisms, such as polymerization, rearrangements of atomic orbital energies, interstitial charge localization, etc..

\textit{Ab-initio} calculations based on Density Functional Theory (DFT) have played a pivotal role in high-pressure research.
Nowadays, these methods permit not only to describe known phases from a microscopic quantum-mechanical viewpoint, but also to predict new structures and properties. The famous Maddox paradox, according to which the quantum mechanical methods for material modelling cannot be considered fully predictive, unless they can  predict crystal structures from the knowledge of the sole chemical composition, has finally been overcome~\cite{Maddox1988}. In fact, modern techniques for crystal structure prediction have  proven their predictive power over a variety of systems, with an astounding agreement with experimental observations  \cite{Oganov2019-wh}.
These techniques utilize clever optimization strategies to identify the global and local minima of the potential energy surface (PES) associated with a given set of atoms,
which 
correspond to the ground-state and metastable 
structures~\cite{Woodley2008}, respectively.
Commonly-employed methods include simulated annealing, \textit{ab-initio} random structure search,~\cite{Pickard_AIRSS_2011} metadynamics,~\cite{Martonak2005-ag}, minima hopping, ~\cite{goedecker2004minima,amsler2010crystal}
evolutionary algorithms,~\cite{glass2006uspex,LONIE2011372,HAJINAZAR2021107679} particle swarm optimization~\cite{Ma_PSO_2010,eberhart1995-xx}, etc..

Indeed, thanks to the increasing integration between experimental and computational methodologies, the knowledge on high-pressure crystal structures has experienced significant advancements in recent years. 
However, the relative information is still largely incomplete, and 
spread over several databases and publications, whose standards vary significantly.
A large portion of these sources is either unaccessible due to paywalls or rely on outdated conventions.
Even for the most basic systems, such as mono-elemental solids, it is frequently challenging to find complete crystal structure information for the entire range of experimentally-accessible pressures, which nowadays exceed 400 GPa.
In fact, particularly at higher pressures,
the only  available 
crystal structures information derives from computational predictions, which significantly differ in terms of breadth and accuracy.

The aim of the 
\emph{High-pressure Elemental Xstals} database (\emph{HEX} database) is to provide a single open-access, 
easily accessible and well-organized database containing the crystal structures
of the first 57 elements of the periodic table (Hydrogen-Lanthanum) at pressures of 0, 100, 200 and 300 GPa.
The database has been constructed compiling all available literature, and comparing with the results
of highly-accurate evolutionary crystal structure prediction calculations \cite{glass2006uspex}, based on plane-wave pseudopotential Density Functional Theory (DFT) total energies.
Our choice to exclude elements beyond lanthanum
is motivated by the need to maintain a consistent 
accuracy throughout the database:
Elements in the lanthanide and actinide series have been excluded,  due to the inadequacy of the pseudopotential approximation for elements with open $f$-shells, while other heavy elements were discarded, because significant spin-orbit interaction may introduce further sources of inaccuracy in the calculations. 
In order to maintain the computational cost manageable, our evolutionary crystal structure prediction runs employ 8-atoms unit cells, and neglect zero-point energy (ZPE) corrections, which should however be negligible for elements beyond the first rows. 

The primary aim of this work is to provide a complete and accurate reference for researchers in various fields. Moreover, by presenting a systematic comparison of high-quality crystal structure prediction results with literature data, the HEX Database also gives an extensive benchmark of the accuracy of crystal structure prediction methods on elemental crystal structures, which nicely complements existing blind tests on molecules. \cite{csp-blind} We find that evolutionary algorithm (EA) predictions reproduce known experimental results in over 95 $\%$ of the cases; most of the 
observed deviations can  be attributed to the use of too small unit cells.

\section*{Methods}\label{Sec:METHODS}

Data contained in the \emph{HEX} database were generated 
by combining literature data with results of evolutionary crystal structure prediction runs. 

We performed a thorough screening of the available literature to identify the ground-state crystal structures of the first 57 elements of the periodic table (H-La), at 0, 100, 200 and 300 GPa. 
Moreover, we performed unconstrained \textit{ab-initio} EA searches for each element and pressure, as explained below.
The structures obtained from the two sources underwent a final relaxation and symmetrization employing the same convergence criteria.  This allowed us to compare the total energies/enthalpies to determine a single ground-state crystal structure for each element and pressure; taken all together, these structures form the first sub-database -- (Database Ground-State).
We also created two other sub-databases, one containing all structures predicted by EA runs -- (Database Evolutionary Algorithm), and the other containing all literature (LIT) structures which turned out to be less energetically favorable than the EA ones (Database Mismatch).
The content and structure of the three sub-databases is described in detail in the Data Records section; here  
 we describe in detail the generation procedure.

\begin{itemize}
\item {\bf EA-generated structures:}
The bulk of our work involved crystal structure prediction runs for the first 57 elements of the periodic table (H-La), 
 over a wide range of pressures.
We employed evolutionary algorithms as implemented in the \textit{Universal Structure Predictor: Evolutionary Xtallography} (USPEX) code\cite{Oganov2006,Oganov2011-jq,LYAKHOV20131172}.
Structural searches for each element were carried out  at 0, 100, 200, 300 GPa for to identify the lowest-enthalpy structure. 
The underlying structural relaxations and total energy calculations are based on Density Functional Theory (DFT), as implemented in the \textit{Vienna Ab Initio Simulation Package} (VASP)\cite{Kresse1996,Kresse1996-le}. We employed Projector Augmented Wave pseudopotentials\cite{Block1994,kresse1999} part of the standard VASP distribution, and Perdew-Burke-Ernzerhof exchange-correlation functional\cite{perdew1996gga}.
For reciprocal $\mathbf{k}$-space integration we used uniform Monkhorst-Pack grids\cite{Monkhorst}  with Methfessel-Paxton smearing\cite{Methfessel} (See Tab. \ref{tab:steps} for further details).

For each combination of element and pressure, we performed  EA searches with an 8-atom unit cell.
The first generations contained 40 structures (Individuals), while each of the following generation contained 20 structures. Each individual was fully relaxed, following  a five-step relaxation procedure with increasing accuracy; the relevant parameters are summarized in Tab. \ref{tab:steps}.
Crystal structure prediction runs lasted for a maximum of 20 generations, and were considered converged when the the lowest-enthalpy structure remained the same for 7 consecutive generations.
Once the evolutionary algorithm search was converged, we collected the ten lowest-enthalpy structures for each element and pressure. These structures underwent a final relaxation, with tighter criteria listed in the  {\em final} row of Tab. \ref{tab:steps}, and finally symmetrized, using  the FINDSYM algorithm by Stokes et al.\cite{Stokes2005-gt, Stokes_site}, with a tolerance criterion of 0.2 \AA.
The lowest-enthalpy structure after symmetrization for each element and pressure 
was selected as the EA ground-state structure. If, after the \textit{final} relaxation and symmetrization, we found more than one structure
to be degenerate in enthalpy within 26 meV (i.e. $k_{B}T$ for $T = 300 K$), we selected the highest-symmetry one.

\item {\bf Literature search:}
We performed a thorough screening of the existing literature
on the crystal structures of the first 57 elements of the periodic table (H-La), at 0, 100, 200 and 300 GPa. 
We chose experimental references rather than theoretical ones, when available,  and more recent papers were selected in favour of older ones. Our bibliographic search was performed as comprehensively as possible using multiple queries and strategies.  However, we cannot rule out that we may have missed some references.

The experimental structures at ambient pressure were extracted from the American Mineralogist Crystal Structure Database\cite{mineralogist}, 
while information on higher pressure was obtained from multiple
sources.

All references, along with the indication on whether they refer to a theoretical or experimental work, are reported in the  \emph{Ref} column of tables \ref{tab:000gs}--\ref{tab:miss}.
Once identified, structures extracted from literature underwent a single run of structural relaxation, with the same settings used for the \textit{final} relaxation of EA-generated structures before their energies were compared with the EA results. The parameters reported in the tables refer to this \textit{final} relaxation.
\end{itemize}

\begin{table*}[!htb]
\centering
\begin{tabular}{|c|c|c|c|c|c|}
\hline 
Step & Plane wave Cutoff & Smearing & k-point Spacing & 
$\Delta E$   &  $\Delta F$ \\
     &   (eV)    &   (eV)   &  ( 2$\pi \times$\AA$^{-1}$) &  (eV) &  (eV/\AA) \\
\hline
  1 & ENMIN & 0.10 & 0.13 & 10$^{-2}$ & 10$^{-1}$ \\
  2 & ENMAX & 0.08 & 0.11 & 10$^{-3}$ & 10$^{-2}$ \\
  3 & 500 & 0.07 & 0.09 &  5$\times$10$^{-4}$ &  5$\times$10$^{-3}$ \\
  4 & 600 & 0.07 & 0.07 &  10$^{-4}$ & 10$^{-3}$ \\
  5 & 600 & 0.06 & 0.04 & 10$^{-5}$ & 10$^{-3}$ \\
  \hline
  final & 600 & 0.06 & 0.04 & 10$^{-5}$ & 10$^{-3}$ \\
\hline
\end{tabular}
\caption{Computational details of the multi-step DFT relaxation
procedure employed in evolutionary crystal structure prediction runs; 
ENMIN/ENMAX indicate the minimum/maximum kinetic energy cutoff values reported in the VASP pseudopotential files; $\Delta E$ and $\Delta F$
indicate the total energy and force convergence criteria, respectively. The final row of the table contains the settings used for the final relaxation of the EA-generated structures as well as literature structures.}
\label{tab:steps}
\end{table*}

\section*{Data Records}
Our \emph{HEX} database comprises the  three sub-databases described below.
Details of the relative structures are reported in the tables \ref{tab:000gs}--\ref{tab:miss}; the corresponding  CIF files can be found on the \textcolor{red}{figshare online repository}.

\begin{itemize}
\item{{\bf DB\_GS (Database Ground-State)}}: The main sub-database includes the ground-state structure for each element at 0, 100, 200, 300 GPa, obtained by comparing the result of our evolutionary crystal structure prediction runs (EA structures) with the structures obtained from the screening of the literature (LIT structures), when available.

 The columns of  Tables \ref{tab:000gs}--\ref{tab:300gs} contain the atomic number \emph{Z}, element symbol, space group, unit cell volume (per atom), and the Wyckoff positions of the ground-state structures; the column \emph{Source} specifies whether the lowest-energy
  structure was found through EA runs (ea), or in literature (lit); an asterisk (*)
  indicates that the EA-generated structure agrees with the literature, while a 
  dash (--) indicates that we could not find a literature reference for the
  relevant element and pressure (\emph{unreported} structures). In cases where the difference in enthalpy between the EA-generated structure and the literature one was below 26 meV/atom,  the structures were considered to be \emph{degenerate}. In the following, structures for which literature and EA results are the same are named \emph{matching}, while those different are named \emph{mismatching}. 
  The column \emph{Ref} reports the literature reference.

\item {{\bf DB\_EA (Database Evolutionary Algorithm)}}: This database contains the results of our evolutionary algorithm searches for every combination of element and pressure considered.
The main results are summarized in tables \ref{tab:000ea}--\ref{tab:300ea}.  The columns contain the atomic number \emph{Z}, element symbol, unit cell volume (per atom), the Wyckoff positions, the relative enthalpy compared to that of the ground-state structure. The relative enthalpy {$\Delta$H$_\text{EA-GS}$} is zero in cases where the EA predicts the lowest-enthalpy structure, and \emph{<26} when the difference between EA and LIT ground-state structure is smaller than 26 meV/atom, and positive otherwise. 
The column \emph{Ref} reports the literature reference.
We indicate in bold-face the entries for which the EA predictions are \emph{unsuccessful}, i.e. cases in which the EA-predicted
structures are neither  \emph{matching} nor \emph{degenerate} with avaiable experimental data.

\item{{\bf DB\_MISS (Database Miss)}}: 
This database contains the list of literature structures less stable than EA-generated ones, and hence not included in the ground-state tables.
The structures for all pressures are grouped into a single table -- Tab.\ref{tab:miss}.
 The columns contain the atomic number \emph{Z}, the element symbol,  the space group when available, the unit cell volume (per atom), the Wyckoff positions, the  enthalpy relative to the ground-state, and the literature reference, together with the indication whether the literature reference is theoretical or experimental. We also indicate explicitely when literature references did not report enough structural information to allow for a comparison with EA-generated structures (\emph{non-reproducible} in the following).
\end{itemize}

\begin{figure}[!htb]
\centering
\includegraphics[width=\linewidth]{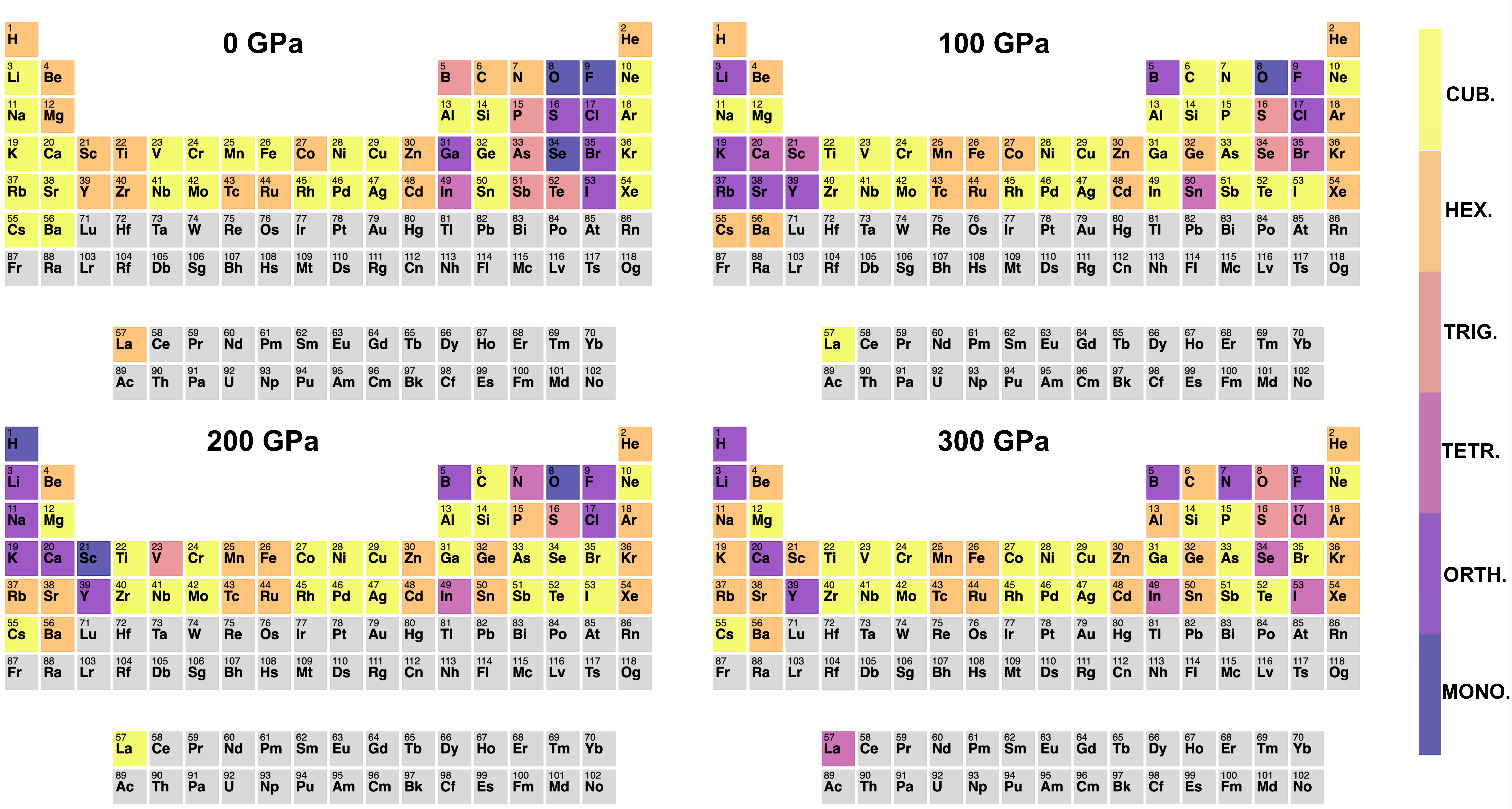}
\caption{Lattice systems for the ground-state structures at 0, 100, 200 and 300 GPa. The colorbar indicates the Bravais lattice: (i) {\em MONO.}: space group 3-15 (monoclinic); (ii) {\em ORTH.} space group 16-74 (orthorombic); (iii) {\em TETR.}  space group  75-142 (tetragonal); (iv) {\em TRIG.}  space group  143-167 (trigonal); (v) {\em HEX.}  space group  168-194 (hexagonal); (vi) {\em CUB.}  space group 195-230 (cubic). The plots have been prepared with the {\em ptable\_trends} program\cite{Rosen_2022}.}
\label{fig:plot}
\end{figure}

In figure \ref{fig:plot} the trends in the evolution of the crystal structure of the elements with pressure are summarized in graphical form.
The four periodic tables indicate, for each element, the lattice system of the ground-state crystal structure at pressures of 0, 100, 200 and 300 GPa: Monoclininc (3-15), Orthorombic(16-74), Tetragonal (75-142), Trigonal (143-167), Hexagonal (168-194), and cubic (195-230). Bravais lattice types are indicated by a color scale, from purple to yellow.

The figure shows that for most elements the evolution of the crystal structure with pressure does not follow the na\"{i}ve expectation that all matter should become more homogeneous under pressure by adopting more close-packed
structures.
In fact, except for transition metals and noble gases,  which adopt either face- or body-centered cubic or hexagonal close-packed structures over the whole range of pressures, other elements undergo a series of transitions, sometimes leading
to very complex structures, which may exhibit lower symmetries than 
ambient-pressure ones. 
The observed deviation from hard-sphere close-packing at high-pressure can originate from different physical mechanisms: charge localization in interstitial sites (electride behavior), in alkali and alkali metals; stabilization of polymeric or molecular phases, in pnictides, chalcogenides and halides; repopulation of atomic orbitals, leading to change in formal valence, as in III and IV-row elements.~\cite{Zhang_materialHP_NRM_2017, Miao2020}

\section*{Technical Validation}
Validation is an intrinsic part of our work, which comprised a thorough  comparison of the results of extensive evolutionary algorithm searches, sampling over 70.000 structures, with available literature data.

\begin{figure}[!htb]
\centering
\includegraphics[width=0.45\columnwidth,angle=0]{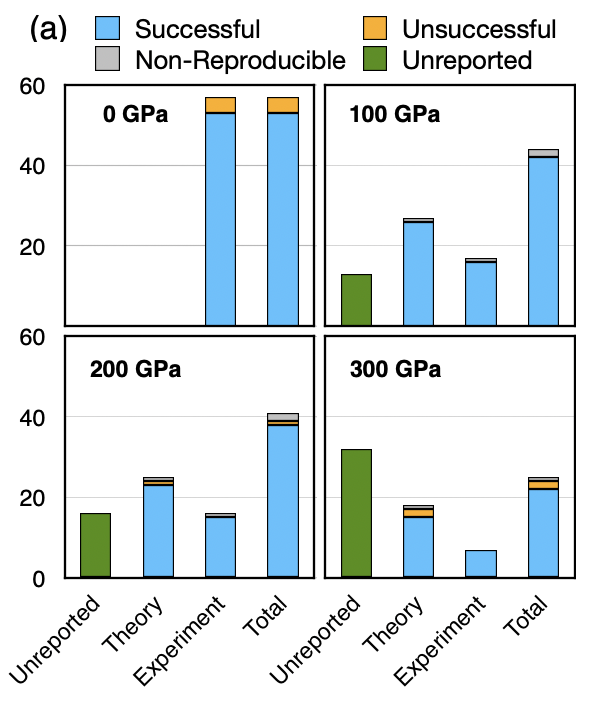}
\includegraphics[width=0.49\columnwidth,angle=0]{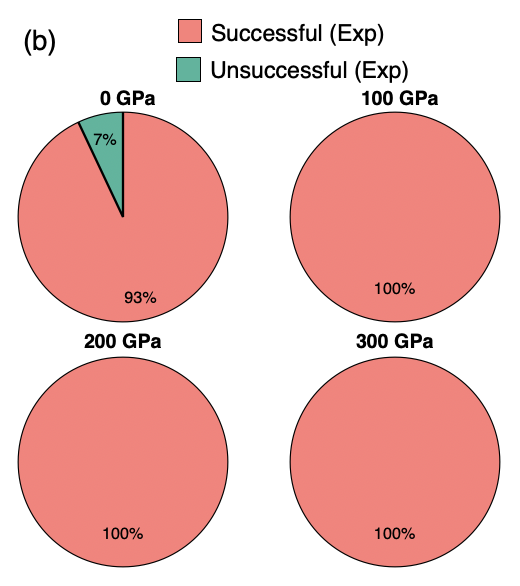}
\caption{(a) The bar chart gives a breakdown of the whole dataset for different pressure into 
structures
(i) Unreported (green) and (ii) reported in literature, divided into \textit{Experiment} and \textit{Theory} category.
The color of the bar indicates whether EA-predicted structures exactly match or are degenerate with available literature
data -- \emph{successful} (blue)/\emph{unsuccessful} (yellow).
(b) Pie chart displaying the fraction of \emph{successful} EA predictions, considering only fully characterized experimental structures.
 Red (green) represents the \emph{successful} (\emph{unsuccessful}) cases for all pressures. 
\label{fig2A:pie}}
\end{figure}

Fig.\ref{fig2A:pie} summarizes the current status of knowledge of high-pressure (HP) structures and presents a comparison with EA-generated structures. 
The bar chart indicates for each pressure
the amount of information available in literature on the structures of the first 57 elements. Structures for each element are divided into \emph{Unreported}, \emph{Theory}, \emph{Experiment}, depending on whether any information is available in literature, and if the source is an experiment or a theoretical
prediction -- The column \emph{Total} is the sum of \emph{Theory} and \emph{Experiment}. The bars are colored to
indicate whether our EA-prediction runs
were \emph{succesful/unsuccesful} in
reproducing literature data. 
A  \emph{succesful} prediction implies that the EA-predicted structure is either exactly \emph{matching} the literature structure
or \emph{degenerate} with it to within 300 K (26 meV).
Cases in which
literature information did not contain enough data to fully reproduce the structures are indicated as \emph{Non-reproducible}.

While at ambient pressure the structures of all these elements  have been experimentally determined and are collected in American Mineralogist Crystal Structure Database\cite{mineralogist}, as pressure increases fewer and fewer experimental reports of high-pressure elemental phases can be found. For example,
at 300 GPa, experimental information 
is available for only about 15$\%$ of the
57 elements considered in this work;
about twice the same amount of structures
can be recovered from theoretical predictions, but for more than 50 $\%$
the structure is \emph{unreported}.

 In general there is a remarkable agreement between our EA predictions and experiment. Moreover, we find that for most cases where we could not identify any literature reference, our EA calculations predict that the elements will retain the same crystal structure measured at lower pressures. 
 In the rare cases in which we observe a disagreement between EA predictions and experiments \textit{a posteriori} it is easy to find very plausible explanations, discussed in the following.

In the right panel of Fig.~\ref{fig2A:pie} we use a pie chart 
to quantify the success rate of EA predictions.
The comparison in this case involves
only cases for which full experimental information is available.
On average, we find that $\sim$ 98 $\%$ of the EA predictions were \emph{succesful},
i.e. EA either predicted the same structure as experiment (\emph{matching} structures),
or a structure \emph{degenerate} with it to within 26 meV. 

\begin{itemize}
\item \textbf{Ambient pressure (0 GPa)}: Of the 57 papers found in literature for 0 GPa, 36 reports are \emph{matching} with our studies. Of the remaining 21 \emph{mismatching} cases, 17 are \emph{degenerate} in enthalpy. 
This means that 53 structures can labeled as \emph{succesful}.

In a few cases, the original 
mismatch between EA predictions and experiment was eliminated including corrections to the standard GGA
functional used for all our calculations.
In particular, for
Br and I, marked with daggers in the tables, the experimental ground-state structures become degenerate in enthalpy with our 
calculated structures after adding Van-Der-Waals corrections.
While experimentally these elements 
form molecular crystals, the structures we predict contain we zig-zag polymeric chains.
Since the two types of structures are almost degenerate in energy, it is conceivable
that, depending on the activation energy and temperature dependence of the polymerization, also polymeric structures might be experimentally realizable.

In order for the EA predictions to match experiments for for Fe, Co and Ni,
we had to include spin polarization in the calculations. These enetries are marked with asterisks in the table.

Of the four \emph{unsuccessful} structures, B, S and Mn have a ground-state characterized by cells much larger than the 8 atoms cell we considered for our EA searches, while for tellurium, we believe that the source of the discrepancy may be a substantial role of spin-orbit effects, which are neglected in our calculations.

In synthesis, at  0 GPa group  93\% of the EA predictions can be defined  \emph{successful}, according to our criteria. 

\item \textbf{100 GPa Group}: Of the 44 structures reported in literature for 100 GPa, 
our EA predictions are \emph{matching}
for 34 elements. Of the remaining 10 \emph{mismatching} cases, 8 are \emph{degenerate} in enthalpy. 
For the remaining two elements, S and In, literature references did not contain enough information to fully reconstruct the structures, only the Bravais lattices -- bco for S\cite{Whaley-Baldwin2020-vs} and bct for In\cite{Simak2000-gx}. Hence, they should be classified as \emph{unreported}.

At 100 GPa, our EA structures are \emph{successful} in reproducing the literature data in 100\% of the cases where complete experimental information was available.

\item \textbf{200 GPa Group}: Of the 41 papers found in literature for 200 GPa,
our EA predictions are \emph{matching}
in 34 cases.
 Of the remaining 7 \emph{mismatching} cases, 4 are \emph{degenerate}.  
We have not been able to gather enough information to perform calculations on the reported phases for N and Sc, which should then be classified as \emph{unreported}.
The EA-predicted structure for Ni (fcc) is more stable than the bcc phase predicted by 
Belashenko et al.~\cite{Belashchenko2020-od}.
Including spin-polarization
in the calculation does not modify this result. It is likely that
 strong correlation effects 
 may solve the discrepancy \cite{Hausoel_NatComm_2017_Ni_corr}.

At 200 GPa,
taking into consideration only fully experimentally-determined structures,  \emph{successful} predictions are hence 100\% of the total.

\item \textbf{300 GPa Group}: Our EA predictions \emph{match} 18
of the 25 structures reported in literature for 300 GPa.  Of the 7 \emph{mismatching} cases, 4 are \emph{degenerate} in enthalpy. 
Of the remaining elements, the reference 
reported for N did not contain 
enough information to fully determine
the crystal structure\cite{Wang2010-ct}, and should
then be considered \emph{unreported}.
 For Li\cite{Lv2011-gt} and Y\cite{Chen2012-xx}, the structures we obtained were found to be less stable than 
 theoretical predictions in literature,
 which however employed much larger unit cells.

In summary, at  300 GPa, our EA structures reproduced  literature results in 92\% of the cases. Taking into consideration only fully-determined experimental structures, the fraction of \emph{successful} predictions rises to 100\%.
\end{itemize}

An exciting outcome of our work is that evolutionary crystal structure predictions based on Density Functional Theory are extremely accurate: on average 96\% of  structures available in literature
were predicted correctly (98\% considering
only fully-determined experimental structures).
In all but two cases where EA-predicted structures could not reproduce the ground-state structures from the literature, we could attribute this either to physical effects not included in our original computational setup (vdW interactions, magnetism, spin-orbit coupling) or to the choice of a too small unit cell. The only two cases for which we could not find a simple explanation are Te at ambient pressure, and Ni at 200 GPa.

\begin{figure}[ht]
\centering
\includegraphics[width=0.6\linewidth]{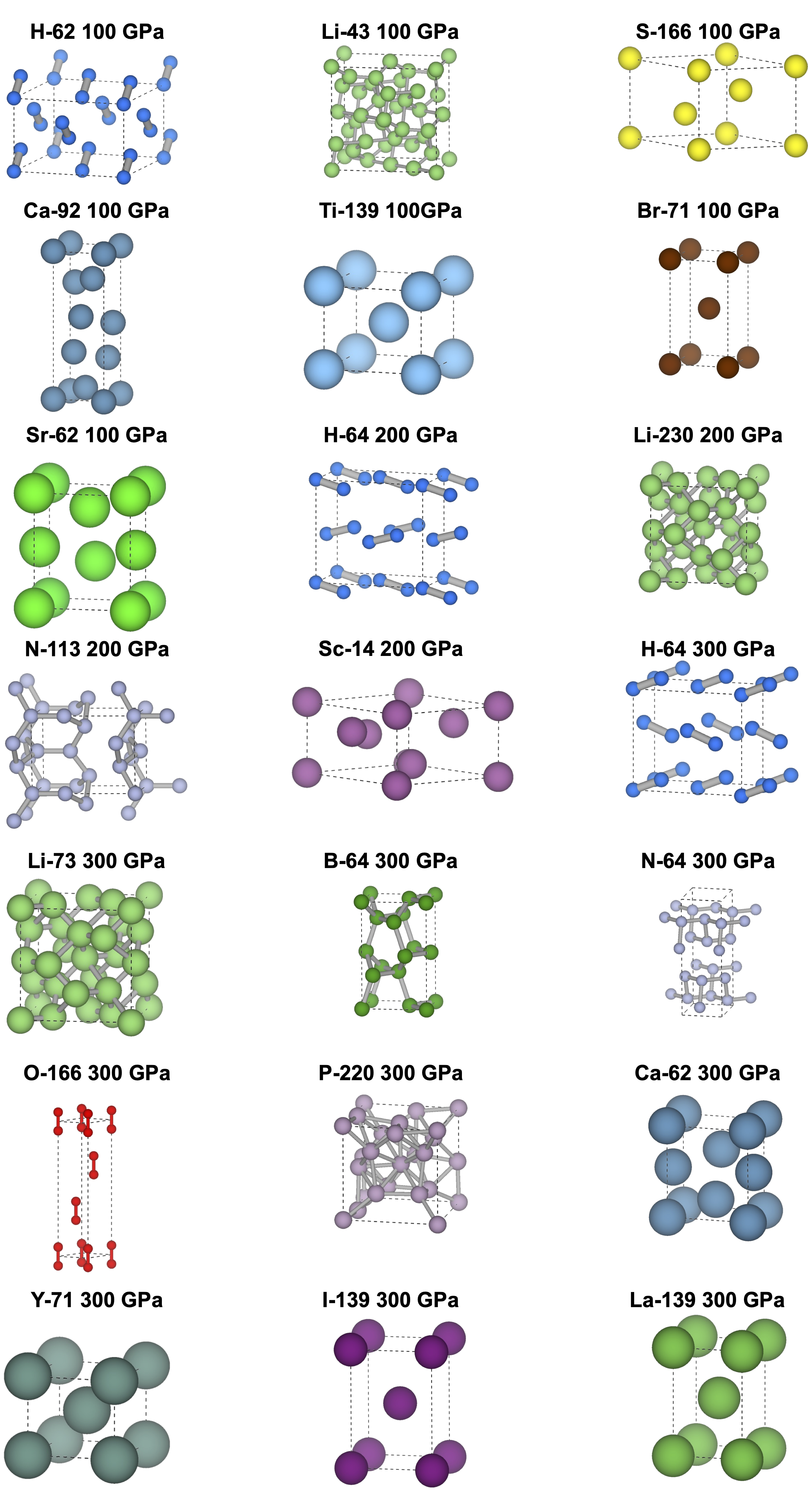}
\caption{EA-predicted crystal structures for elements and pressure where the experimental information is either completely missing,
or too incomplete to reconstruct the structure. We leave out trivial cases in which the structure is a monoatomic \emph{fcc}, \emph{bcc} or \emph{hcp} one.
 Structures are labelled as: Element-Space Group number and pressure.}
\label{fig:crystals}
\end{figure}

In Fig.\ref{fig:crystals}, we show EA-generated crystal structures
for the 21 cases which we believe may be of interest for future studies, labeled with their element, space group number, and pressure.  
In all these cases, experimental information is either not  available at all,
 or too incomplete to completely determine the structure.
 We decided to not show, however, trivial cases in which EA predicted  is a monoatomic \emph{bcc}, \emph{fcc} or \emph{hcp}
 ground-state structures. 

Of the structures shown in figure, hydrogen and oxygen tend to form such strong bonds,
 that they form molecular crystals up to the 
 the highest pressure considered in this work.
 Nitrogen and boron, whose covalent bonds are more prone to frustration, form complex crystalline polymers. Lithium and phosphorous form complex, high-symmetry phases with large unit cells. Heavier elements tend to form less exotic structures, mainly tetragonal distorsions of cubic structures. We note that the qualitative behavior is consistent with what is observed in other elements, where high-pressure data is available.

\section*{Usage Notes}
Data are stored on \textcolor{red}{figshare online repository},
in two separate compressed zip archives.
The first archive - \emph{HEX.zip} - contains three folders,
one for each of the databases described in the text (GS, EA, MISS). Moreover, each folder contains four sub-folders,
one for each pressure.
  The sub-folders contain  files in the standard Crystallographic Information File (cif), named as \emph{ELEMENT\_PRESSURE\_DATABASE.cif} (\emph{DATABASE= GS, EA, MISS}). 
  The second archive -- \emph{Evolutionary.zip} -- contain
  the input files used for the evolutionary prediction runs (USPEX input files + example of VASP INCAR files).

\section*{Code availability}

All calculations described in the paper have been carried out using the \emph{Vienna ab-initio Simulation Package} (VASP), v 6.3.0, for DFT total energies, forces, and structural relaxations, the \emph{Universal Crystal 
Structure Predictor} (USPEX), v 10.5, for crystal structure
searches. The methods section contains all details needed to
reproduce the calculations.
\bibliography{references}

\section*{Acknowledgements} 
L.B., S.D.C. and S.S. acknowledge funding from the Austrian Science Fund (FWF) project number P30269-N36. L.B. and S.D.C. acknowledge support from Fondo Ateneo-Sapienza 2018-2022. F.G. and S.D.C. acknowledge computational resources from CINECA, proj. IsC90-HTS-TECH and IsC99-ACME-C. L.B. and S.d.C. acknowledge support from Project PE0000021, “Network 4 Energy Sustainable Transition – NEST”, funded  by the European Union – NextGenerationEU, under the National Recovery and Resilience Plan (NRRP), Mission 4 Component 2 Investment 1.3 - Call for tender No. 1561 of 11.10.2022 of Ministero dell’Universit\'{a} e della Ricerca (MUR).

\section*{Author contributions statement}
F.G. and L.B. wrote the main draft. F.G. performed the literature search. F.G., S.S. and S.D.C. prepared figures and tables.
F.G. and S.D.C. performed the structural prediction runs and relaxations. All authors participated in the discussions and revised the manuscript.

\section*{Competing interests} 

The authors declare no competing interest.

\section*{Figures \& Tables}

\newpage


\begin{table}[ht]
\centering
\fontsize{6.5}{8}\selectfont
\begin{tabular}{l l l l >{\em}l l l}
\toprule
 Z & Element &  Space group &  Volume &  \normalfont{Wyckoff positions} & Source & Ref \\
 & & & (\AA$^3/atom$) & \\
\midrule
 1 &       H &      194 &                   14.77 &      (4f) z=-0.17853 & lit. & exp.\cite{Vindryavskyi1980-cn} \\
 2 &      He &          194 &                   17.30 &      (2c) &  * &    exp.\cite{Henshaw1958-wj} \\
 3 &      Li &          229 &                   21.27 &      (2a) & lit. &     exp.\cite{Barrett1956-sw} \\
 4 &      Be &          194 &                   7.91 &      (2c) & * &      exp.\cite{Larsen1984-hg} \\
 5 &       B &          166 &                   7.26 &      (18h) x=-0.21455   z=0.22487 & lit. &    exp.\cite{Decker1959-qb} \\
 & & & & (18h) x=0.19695 z=0.02424  & \\
 6 &       C &          191 &                   10.34 &      (2d) & * &       exp.\cite{Zemann1965-mf} \\
 7 &       N &  194 &                  33.54 &    (4f) z=-0.33949 & lit. &       exp.\cite{Schuch1970-tj} \\
 8 &       O &    12 &                   14.58 &      (4i) x=-0.14455 z=0.12053 & lit. &       exp.\cite{Datchi2014-pf} \\
 & & & & (4i) x=0.20159 z=0.18626  & \\
 & & & & (8j) x=0.02857 y=0.23724 z=0.15337  & \\
 9 &       F &           12 &                  17.90 &      (4i) x=0.05400 z=0.10073 & lit. &        exp.\cite{Meyer1968-on} \\
 & & & & (4i) x=-0.49221 z=-0.39902  & \\
10 &      Ne &          225 &                   22.36 &      (4a) & * &       exp.\cite{Zemann1965-mf} \\
11 &      Na &          229 &                   37.71 &      (2a) & lit. &       exp.\cite{Zemann1965-mf} \\
12 &      Mg &          194 &                   23.09 &      (2d) & * &       exp.\cite{Zemann1965-mf} \\
13 &      Al &          225 &                   16.53 &      (4a) & * &       exp.\cite{Zemann1965-mf} \\
14 &      Si &          227 &                   20.46 &      (8b) & * &      exp.\cite{Zemann1965-mf} \\
15 &       P &          164 &                   23.84 &      (2d) z=0.12144 & * &        exp.\cite{Cartz1979-wd} \\
16 &       S &         70   &              34.54 &  (32h) x=0.20642 y=0.05407 z=0.10800 & lit. &      exp.\cite{Rettig1987-em} \\
& & & & (32h) x=-0.24429 y=0.02723 z=0.47620  & \\
& & & & (32h) x=-0.17438 y=0.48401 z=0.03896  & \\
& & & & (32h) x=0.37513 y=0.09684 z=0.04511 & \\
17 &      Cl &           64 &                   34.21 &      (8f) y=-0.11492 z=-0.10573 & lit. &       exp.\cite{Powell1984-tf} \\
18 &      Ar &          225 &                   45.84 &      (4a) & * &       exp.\cite{Zemann1965-mf} \\
19 &       K &          229 &                   73.39 &      (2a) & lit. &       exp.\cite{Zemann1965-mf} \\
20 &      Ca &          225 &                   42.58 &      (4a) & * &       exp.\cite{Zemann1965-mf} \\
21 &      Sc &          194 &                   24.52 &      (2d) & * &       exp.\cite{Zemann1965-mf} \\
22 &      Ti &          194 &                   17.42 &      (2c) & * &       exp.\cite{Zemann1965-mf} \\
23 &       V &          229 &                   13.50 &      (2a) & * &        exp.\cite{Zemann1965-mf} \\
24 &      Cr &          229 &                   11.46 &      (2a) & * &       exp.\cite{Zemann1965-mf} \\
25 &      Mn &          217 &                   10.74 &      (2a) & lit. &  exp.\cite{Oberteuffer1970-ll} \\
& & & & (8c) x=0.18194  & \\
& & & & (24g) x=0.14325 z=0.46199  & \\
& & & & (24g) x=0.41117 z=0.21894  & \\
26 &      Fe* &          229 &                   11.77 &      (2a) & * &         exp.\cite{Wilburn1978} \\
27 &      Co* &          194 &                   10.20 &      (2c) & * &       exp.\cite{Zemann1965-mf} \\
28 &      Ni* &          225 &                   10.76 &      (4a) & * &       exp.\cite{Zemann1965-mf} \\
29 &      Cu &          225 &                   12.00 &      (4a) & * &       exp.\cite{Zemann1965-mf} \\
30 &      Zn &          194 &                   15.25 &      (2d) & * &       exp.\cite{Zemann1965-mf} \\
31 &      Ga &           64 &                   20.53 &      (8f) y=0.34222 z=0.41786 & * &       exp.\cite{Sharma1962-sm} \\
32 &      Ge &          227 &                   24.15 &      (8a) & * &          exp.\cite{Hom1975-bu} \\
33 &      As &          166 &                   22.93 &      (6c) & * &     exp.\cite{Schiferl1969-ku} \\
34 &      Se &            14 &                   39.00 & (4e) x=0.06913 y=0.49655 z=0.24516 & lit. &      exp.\cite{Cherin1972-ej} \\
& & & & (4e) x=0.05854 y=0.66250 z=0.35690  & \\
& & & & (4e) x=-0.21494 y=-0.36324 z=-0.46970  & \\
& & & & (4e) x=-0.40449 y=-0.20629 z=-0.45721  & \\
& & & & (4e) x=0.40877 y=-0.30798 z=-0.48887  & \\
& & & & (4e) x=-0.47828 y=-0.27649 z=0.32991  & \\
& & & & (4e) x=-0.31228 y=-0.45720 z=0.23086  & \\
& & & & (4e) x=-0.01374 y=-0.40631 z=0.14476  & \\
35 &      Br\textsuperscript{\textdagger} &         64 & 30.88 & (8f) y=0.15416 z=0.11932 & lit. &  exp.\cite{Powell1984-tf} \\
36 &      Kr &          225 &                   66.44 &      (4a) & * &       exp.\cite{Zemann1965-mf} \\
37 &      Rb &          229 &                   91.09 &      (2a) & lit. &       exp.\cite{Zemann1965-mf} \\
38 &      Sr &          225 &                   54.97 &      (4a) & * &       exp.\cite{Zemann1965-mf} \\
39 &       Y &          194 &                   32.53 &      (2c) & * &       exp.\cite{Zemann1965-mf} \\
40 &      Zr &          194 &                   23.44 &      (2d) & * &       exp.\cite{Zemann1965-mf} \\
41 &      Nb &          229 &                   18.16 &      (2a) & * &       exp.\cite{Zemann1965-mf} \\
42 &      Mo &          229 &                   15.83 &      (2a) & * &       exp.\cite{Zemann1965-mf} \\
43 &      Tc &          194 &                   14.21 &      (2c) & lit. & exp.\cite{Zemann1965-mf} \\
44 &      Ru &          194 &                   13.71 &      (2c) & * &       exp.\cite{Zemann1965-mf} \\
45 &      Rh &          225 &                   14.08 &      (4a) & * &       exp.\cite{Zemann1965-mf} \\
46 &      Pd &          225 &                   15.30 &      (4a) & * &       exp.\cite{Zemann1965-mf} \\
47 &      Ag &          225 &                   18.00 &      (4a) & * &          exp.\cite{Suh1988-rz} \\
48 &      Cd &          194 &                   22.79 &      (2c) & lit. &       exp.\cite{Zemann1965-mf} \\
49 &      In &          139 &                   27.43 &      (2a) & lit. &        exp.\cite{Smith1964-gm} \\
50 &      Sn &          227 &                   36.87 &      (8a) & * &        exp.\cite{Smith1964-gm} \\
51 &      Sb &          166 &                   32.13 &      (6c) z=0.26654 & * &     exp.\cite{Schiferl1969-ku} \\
52 &      Te &          152 &                   34.99 &      (3a) x=0.73640 & lit. &       exp.\cite{Adenis1989-jc} \\
53 &       I$^\dag$ &           64 & 41.31 & (8f) y=-0.16897 z=0.37788 & lit. &  exp.\cite{Zemann1965-mf} \\
54 &      Xe &          225 &                   88.43 &      (4a) & * &       exp.\cite{Zemann1965-mf} \\
55 &      Cs &          229 &                   117.16 &      (2a) & lit. &       exp.\cite{Zemann1965-mf} \\
56 &      Ba &          229 &                   64.07 &      (2a) & * &       exp.\cite{Zemann1965-mf} \\
57 &      La &          194 &                   37.56 &      (2a) & lit. &       exp.\cite{Zemann1965-mf} \\
& & & & (2d) & \\
\bottomrule
\end{tabular}
\caption{\label{tab:000gs}Ground-state database (DB\_GS) at 0 GPa. The symbols in the \textit{Source} column indicate: 
(i) \emph{*}  the structure is \emph{matching}; (ii) \emph{-}  no  reference could be found in literature; (iii) \emph{lit.}/\emph{ea.} the 
ground-state 
structure originates from literature/evolutionary algorithm. In the 
\emph{Ref} column, (i) \emph{th.} (\emph{exp.}) specifies whether the literature source is computational (experimental), 
or (ii) \emph{-}  missing; (iii) \emph{Fe*}, \emph{Co*} and \emph{Ni*} indicate that
the calculation is spin-polarized
with a magnetic moment of 2.22, 1.74, 0.606 $a.u.$ respectively\cite{Galperin1978-dz} and (iv) {$Br^\dag$} and {$I^\dag$}, 
indicate that the calculation employed the opt88-vdW \textit{xc} functional \cite{vdw_michaelides_2010}.}
\end{table}


\begin{table}[ht]
\centering
\fontsize{7}{8.5}\selectfont
\begin{tabular}{l l l l >{\em}l l l}
\toprule
 Z & Element &  Space group &  Volume &  \normalfont{Wyckoff positions} & Source & Ref \\
 & & & (\AA$^3/atom$) & \\
\midrule
1 &       H &          176 &         2.31 &     (4f) z=-0.37200 & lit. &                              th.\cite{Pickard2007-lq} \\
& & & & (6h) x=0.09840 y=0.39006  & \\
& & & & (6h) x=0.19824 y=0.26598  & \\
 2 &      He &          194 &                   3.46 &      (2c) & - &                                                 - \\
 3 &      Li &     64 &                  6.00 &      (8f) y=-0.17426 z=0.43844 & lit. &                                  th.\cite{Lv2011-gt} \\
 & & & & (16g) x=0.33351 y=0.10864 z=0.34339  & \\
 4 &      Be &          194 &                   5.27 &      (2c) & * &                                th.\cite{Kadas2007-ek} \\
 5 &       B &           64 &                   5.00 &      (8f) y=-0.15747 z=-0.08761 & * &          exp.\cite{Shirai2011-xb} \\
 6 &       C &          227 &                   4.83 &      (8b) & * &                      th.\cite{Grumbach1996-ak} \\
 7 &       N &          199 &                   5.52 &      (8a) x=0.32248 & * &                                 th.\cite{Wang2010-ct} \\
 8 &       O &           15 &                   6.00 &      (8f) x=0.09795 y=0.29000 z=0.21503  & * &                                   th.\cite{Ma2007-vj} \\
 9 &       F &           64 &                   6.29 &      (8f) y=0.34595 z=0.11698 & * &                                th.\cite{Olson2020-ng} \\
10 &      Ne &          225 &                   6.71 &      (4a) & * &                              exp.\cite{Dewaele2008-qe} \\
11 &      Na &          225 &                   10.78 &      (4a) & * &                             exp.\cite{Hanfland2002-oc} \\
12 &      Mg &          229 &                   11.25 &      (2a) & * &                                   th.\cite{Li2010-hy} \\
13 &      Al &          225 &                   10.34 &      (4a) & * &                               exp.\cite{Fiquet2019-wc} \\
14 &      Si &          225 &                   9.19 &      (4a) & * &                               exp.\cite{Mujica2003-nr} \\
15 &       P &          221 &                   9.95 &      (1a) & * &                              exp.\cite{Akahama1999-yg} \\
16 &       S &          166 &                   9.99 &      (3b) & ea. &                              exp.\cite{Whaley-Baldwin2020-vs} \\
17 &      Cl &           64 &                   11.40 &      (8f) y=0.31818 z=-0.37903 & * &                                th.\cite{Olson2020-ng} \\
18 &      Ar &          194 &                   12.62 &      (2d) & - &                                                  - \\
19 &       K &           64 &                   11.81 &      (8d) x=-0.28496 & * &                                   th.\cite{Ma2008-yy} \\
& & & & (8f) y=0.32494 z=0.32775  & \\
20 &      Ca &           96 &                   12.44 &      (8b) x=-0.01497 y=0.32129 z=-0.09575 & lit. &             th.\cite{Oganov2010-my} \\
21 &      Sc &           88 &                   12.11 &      (16f) z=-0.27743 y=-0.19917 z=0.47208 & * &                 exp.\cite{Akahama2005-yi} \\
22 &      Ti &          229 &                   10.77 &      (2a) & lit. &                              th.\cite{Kutepov2003-jg} \\
23 &       V &          229 &                   9.85 &      (2a) & * &                                th.\cite{Verma2008-on} \\
24 &      Cr &          229 &                   9.06 &      (2a) & - &                                                  - \\
25 &      Mn &          194 &                   8.48 &      (2c) & * &                          exp.\cite{Magad-Weiss2020-pd} \\
26 &      Fe &          194 &                   8.18 &      (2c) & * &                                  exp.\cite{Mao1990-rd} \\
27 &      Co &          194 &                   8.15 &      (2c) & lit. &                                  exp.\cite{Yoo2000-dz} \\
28 &      Ni &          225 &                   8.31 &      (4a) & * &                         th.\cite{Belashchenko2020-od} \\
29 &      Cu &          225 &                   8.67 &      (4a) & - &                                                  - \\
30 &      Zn &          194 &                   9.64 &      (2c) & * &                              exp.\cite{Akahama2021-tw} \\
31 &      Ga &          225 &                   10.93 &      (4a) & * &                              th.\cite{Simak2000-gx} \\
32 &      Ge &          191 &                   11.88 &      (1a) & lit. &                              exp.\cite{Akahama2021-tw} \\
33 &      As &          229 &                   12.02 &      (2a) & * &                                th.\cite{Silas2008-af} \\
34 &      Se &          166 &                   12.72 &      (3b) & * &                           exp.\cite{Degtyareva2005-bf} \\
35 &      Br &           139 &             23.08 &      (2a) & lit. &                                  th.\cite{Li2020-pg} \\
36 &      Kr &          194 &                   15.58 &      (2c) & - &                                                  - \\
37 &      Rb &           64 &                   15.00 &      (8d) x=0.21706 & * &                           th.\cite{Ma2008-yy} \\
& & & & (8f) y=-0.32320 z=-0.17520  & \\
38 &      Sr &           62 &                   14.96 &      (4c) x=0.17310 z=-0.07136 &  - &                                                  - \\
39 &       Y &           70 &                   14.66 &      (16g) z=0.43748 &  lit. &  exp.\cite{Pace2020-ih} \\
40 &      Zr &          229 &                   13.82 &      (2a) &  * &                           exp.\cite{Anzellini2020-uc} \\
41 &      Nb &          229 &                   13.00 &      (2a) &  * &                          th.\cite{Krasilnikov2014-pn} \\
42 &      Mo &          229 &                   12.48 &      (2a) &  * &                          th.\cite{Krasilnikov2014-pn} \\
43 &      Tc &          194 &                   11.73 &      (2d) &  * &                                 th.\cite{Shah2021-ul} \\
44 &      Ru &          194 &                   11.26 &      (2c) &  * &                                  th.\cite{Liu2020-bg} \\
45 &      Rh &          225 &                   11.26 &      (4a) &  - &                                                  - \\
46 &      Pd &          225 &                   11.61 &      (4a) &  * &                                    th.\cite{National_University_of_Defense_Technology_Changsha2021-nk} \\
47 &      Ag &          225 &                   12.28 &      (4a) &  - &                                                  - \\
48 &      Cd &          194 &                   13.52 &      (2d) &  - &                                                  - \\
49 &      In &          225 &                   15.07 &      (4a) &  ea. &                                th.\cite{Simak2000-gx} \\
50 &      Sn &          139 &                   15.89 &      (2b) &  * &                                   th.\cite{Yu2006-cz} \\
51 &      Sb &          229 &                   16.39 &      (2a) &  - &                                                  - \\
52 &      Te &          225 &                   16.53 &      (4a) &  * &                                  th.\cite{Liu2018-gy} \\
53 &       I &          225 &                   17.35 &      (4a) &  - &                                                  - \\
54 &      Xe &          194 &                   20.05 &      (2c) &  - &                                                  - \\
55 &      Cs &          194 &                   19.00 &      (2a) &  * &                                 th.\cite{Guan2020-lt} \\
& & & & (2c)  & \\
56 &      Ba &          194 &                   18.73 &      (2d) & - &                                                  - \\
57 &      La &          225 &                   16.73 &      (4a) & * &                                 exp.\cite{Chen2022-wm} \\
\bottomrule
\end{tabular}
\caption{\label{tab:100gs}Ground-state database (DB\_GS) at 100 GPa. The symbols in the \textit{Source} column indicate: 
(i) \emph{*}  the structure is \emph{matching}; (ii) \emph{-}  no  reference could be found in literature; (iii) \emph{lit.}/\emph{ea.} the 
ground-state 
structure originates from literature/evolutionary algorithm. In the 
\emph{Ref} column, (i) \emph{th.} (\emph{exp.}) specifies whether the literature source is computational (experimental), or
(ii) \emph{-}  missing.}
\end{table}


\begin{table}[ht]
\centering
\fontsize{7}{8.5}\selectfont
\begin{tabular}{l l l l >{\em}l l l}
\toprule
 Z & Element &  Space group &  Volume &  \normalfont{Wyckoff positions} & Source & Ref \\
 & & & (\AA$^3/atom$) & \\
\midrule
 1 &       H &     15 &       1.73 &      (8f) x=0.26735 y=0.42295 z=0.24415 & lit. &      th.\cite{Pickard2007-lq} \\
 & & & & (8f) x=0.15706 y=0.30406 z=0.22058   & \\
 & & & & (4e) y=-0.35218   & \\
 & & & & (4e) y=-0.10217   & \\
 2 &      He &          194 &                   2.70 &      (2c) & - &                          - \\
 3 &      Li &          64 &        4.48 &   (8f) y=-0.44022 z=-0.33050 & lit. &           th.\cite{Lv2011-gt} \\
 & & & & (16g) x=0.14346 y=0.16867 z=0.06475   & \\
 & & & & (16g) x=-0.28580 y=0.39461 z=0.34794   & \\
 & & & & (16g) x=-0.07142 y=0.13770 z=-0.37225   & \\
 4 &      Be &          194 &                   4.33 &      (2d) & * &       th.\cite{Kadas2007-ek} \\
 5 &       B &           64 &                   4.34 &      (8f) y=0.34220 z=0.41304 & * &       exp.\cite{Shirai2011-xb} \\
 6 &       C &          227 &                   4.34 &      (8b) & * &     th.\cite{Grumbach1996-ak} \\
 7 &       N &          113 &                   4.64 &      (4e) x=-0.33587 z=0.32228 & ea. &         th.\cite{Wang2010-ct} \\
 & & & & (4d) z=-0.16015   & \\
 8 &       O &           12 &                   5.08 &      (4i) x=0.22601 z=-0.20344 & * &           th.\cite{Ma2007-vj} \\
 & & & & (4i) x=-0.29538 z=-0.20316   & \\
 & & & & (8j) x=-0.46534 y=0.25994 z=0.20327  & \\
 9 &       F &           64 &                   5.28 &      (8f) y=-0.16484 z=-0.11903 & * &        th.\cite{Olson2020-ng} \\
10 &      Ne &          225 &                   5.54 &      (4a) & * &      exp.\cite{Dewaele2008-qe} \\
11 &     Na &           62 &                   8.00 &      (4c) x=0.32391 z=-0.08020 & ea. &           exp.\cite{Ma2009-ey} \\
& & & & (4c) x=0.48576 z=0.30971   & \\
12 &      Mg &          229 &                   9.02 &      (2a) & * &           th.\cite{Li2010-hy} \\
13 &      Al &          225 &                   8.62 &      (4a) & lit. &       exp.\cite{Fiquet2019-wc} \\
14 &      Si &          225 &                   7.81 &      (4a) & * &       exp.\cite{Mujica2003-nr} \\
15 &       P &          191 &                   7.90 &      (1b) & * &      exp.\cite{Akahama2000-sg} \\
16 &       S &          166 &                   8.13 &      (3b) & * &          exp.\cite{Luo1993-pi} \\
17 &      Cl &           71 &                   9.07 &      (2a) & * &        th.\cite{Olson2020-ng} \\
18 &      Ar &          194 &                   10.20 &     (2c) & - &                          - \\
19 &       K &           64 &                   9.73 &      (8d) x=-0.21662 & * &           th.\cite{Ma2008-yy} \\
& & & & (8f) y=0.32384 z=0.17522  & \\
20 &      Ca &           62 &                   9.30 &      (4c) x=0.33247 z=-0.39914 & * &       exp.\cite{Sakata2011-qz} \\
21 &      Sc &      14 &       9.22 &     (4e) x=-0.24789 y=-0.08441 z=0.29798 & ea. &      exp.\cite{Akahama2005-yi} \\
22 &      Ti &          229 &                   8.72 &      (2a) & * &      th.\cite{Kutepov2003-jg} \\
23 &       V &          166 &                   8.39 &      (3a) & * &        th.\cite{Verma2008-on} \\
24 &      Cr &          229 &                   7.97 &      (2a) & - &                          - \\
25 &      Mn &          194 &                   7.57 &      (2d) & * &  exp.\cite{Magad-Weiss2020-pd} \\
26 &      Fe &          194 &                   7.29 &      (2c) & * &          exp.\cite{Mao1990-rd} \\
27 &      Co &          225 &                   7.24 &      (4a) & * &          exp.\cite{Yoo2000-dz} \\
28 &      Ni &          225 &              7.34 &      (4a) & ea. &    th.\cite{Belashchenko2020-od} \\
29 &      Cu &          225 &                   7.57 &      (4a) & - &                          - \\
30 &      Zn &          194 &                   8.28 &      (2c) & * &      exp.\cite{Akahama2021-tw} \\
31 &      Ga &          225 &                   9.21 &      (4a) & * &        th.\cite{Simak2000-gx} \\
32 &      Ge &          194 &                   9.80 &      (2c) & * &      exp.\cite{Akahama2021-tw} \\
33 &      As &          229 &                   10.15 &      (2a) & * &        th.\cite{Silas2008-af} \\
34 &      Se &          229 &                   10.62 &      (2a) & * &      exp.\cite{Akahama2021-tw} \\
35 &      Br &          225 &                   11.20 &      (4a) & * &           th.\cite{Li2020-pg} \\
36 &      Kr &          194 &                   12.63 &      (2d) & - &                          - \\
37 &      Rb &          194 &                   12.18 &      (2a) & * &           th.\cite{Ma2008-yy} \\
& & & & (2d)  & \\
38 &      Sr &          194 &                   11.92 &      (2d) & - &                          - \\
39 &       Y &           70 &                   11.64 &      (16e) x=0.43748 & * &       th.\cite{LI2019-x} \\
40 &      Zr &          229 &                   11.32 &      (2a) & - &                          - \\
41 &      Nb &          229 &                   11.17 &      (2a) & * &  th.\cite{Krasilnikov2014-pn} \\
42 &      Mo &          229 &                   10.97 &      (2a) & * &  th.\cite{Krasilnikov2014-pn} \\
43 &      Tc &          194 &                   10.35 &      (2c) & - &                          - \\
44 &      Ru &          194 &                   10.09 &      (2c) & * &          th.\cite{Liu2020-bg} \\
45 &      Rh &          225 &                   10.06 &      (4a) & - &                          - \\
46 &      Pd &          225 &                   10.25 &      (4a) & - &                          - \\
47 &      Ag &          225 &                   10.69 &      (4a) & - &                          - \\
48 &      Cd &          194 &                   11.50 &      (2c) & - &                          - \\
49 &      In &          139 &                   12.68 &      (2b) & * &        th.\cite{Simak2000-gx} \\
50 &      Sn &          194 &                   13.40 &      (2c) & * &           th.\cite{Yu2006-cz} \\
51 &      Sb &          229 &                   13.80 &      (2a) & - &                          - \\
52 &      Te &          225 &                   13.96 &      (4a) & * &          th.\cite{Liu2018-gy} \\
53 &       I &          225 &                   14.37 &      (4a) & - &                          - \\
54 &      Xe &          194 &                   16.24 &      (2c) & - &                          - \\
55 &      Cs &          225 &                   15.60 &      (4a) & * &         th.\cite{Guan2020-lt} \\
56 &      Ba &          194 &                   15.67 &      (2d) & - &                          - \\
57 &      La &          225 &                   14.14 &      (4a) & * &        exp.\cite{Chen2022-wm} \\
\bottomrule
\end{tabular}
\caption{\label{tab:200gs}Ground-state database (DB\_GS) at 200 GPa. The symbols in the \textit{Source} column indicate: 
(i) \emph{*}  the structure is \emph{matching}; (ii) \emph{-}  no  reference could be found in literature; (iii) \emph{lit.}/\emph{ea.}  the ground-state 
structure originates from literature/evolutionary algorithm. In the 
\emph{Ref} column, (i) \emph{th.} (\emph{exp.}) specifies whether the literature source is computational (experimental), or
(ii) \emph{-}  missing.}
\end{table}


\begin{table}[ht]
\centering
\fontsize{7}{8.5}\selectfont
\begin{tabular}{l l l l >{\em}l l l}
\toprule
 Z & Element &  Space group &  Volume &  \normalfont{Wyckoff positions} & Source &         Ref \\
 & & & (\AA$^3/atom$) & \\
\midrule
 1 &       H &      64 &     1.45 &  (8f) x=0.00366 z=0.13499 & lit. &    th.\cite{Pickard2007-lq} \\
 & & & & (4f) x=0.13176 z=0.45360  & \\
 & & & & (4f) x=0.26798 z=0.31709  & \\
 2 &      He &          194 &                   2.32 &       (2d) & - &                         - \\
 3 &      Li &           64 & 3.80 &  (8f) y=0.05146 z=0.17196 & lit. &         th.\cite{Lv2011-gt} \\
 & & & & (16g) x=-0.35620 y=0.17042 z=-0.44172  & \\
 & & & & (16g) x=0.21371 y=0.39288 z=-0.15060  & \\
 & & & & (16g) x=0.42870 y=0.13685 z=0.12675  & \\
 4 &      Be &          194 &                   3.81 &       (2c) & * &          th.\cite{Lu2015-aa} \\
 5 &       B &           64 &                   3.93 &       (8f) y=0.157750 z=-0.08527 & - &                         - \\
 6 &       C &          194 &                   4.00 &       (4e) z=-0.09278 & - &                         - \\
  & & & & (4f) z=-0.34395  & \\
 7 &       N &           64 &                   4.18 &       (8f) y= 0.10538 z=-0.10004 & ea. &        th.\cite{Wang2010-ct} \\
 8 &       O &          166 &                   4.55 &       (6c) z=-0.06915 & - &                         - \\
 9 &       F &           64 &                   4.72 &       (8f) y=-0.32735 z=-0.37974 & * &       th.\cite{Olson2020-mk} \\
10 &      Ne &          225 &                   4.91 &       (4a) & - &                         - \\
11 &      Na &          194 &                   6.68 &       (2a) & * &         exp.\cite{Ma2009-ey} \\
& & & & (2d)  & \\
12 &      Mg &          229 &                   7.84 &       (2a) & * &          th.\cite{Li2010-hy} \\
13 &      Al &          194 &                   7.57 &       (2c) & * &     exp.\cite{Akahama2006-gw} \\
14 &      Si &          225 &                   7.00 &       (4a) & - &                         - \\
15 &       P &          220 &                   6.84 &       (16c) x=0.47267 & - &                         - \\
16 &       S &          166 &                   7.15 &       (3a) & * &  th.\cite{Degtyareva2007-mq} \\
17 &      Cl &          139 &                   7.90 &       (2a) & * &       th.\cite{Olson2020-ng} \\
18 &      Ar &          194 &                   8.92 &       (2c) & - &                         - \\
19 &       K &          194 &                   8.47 &       (2a) & * &          th.\cite{Ma2008-yy} \\
& & & & (2d)  & \\
20 &      Ca &           62 &                   8.07 &       (4c) & - &      - \\
21 &      Sc & 180 &  7.83 &  (3c) & lit. &     exp.\cite{Akahama2005-yi} \\
22 &      Ti &          229 &                   7.61 &       (2a) & * &       exp.\cite{Zhang2022-jj} \\
23 &       V &          229 &                   7.57 &       (2a) & * &       th.\cite{Verma2008-on} \\
24 &      Cr &          229 &                   7.28 &       (2a) & - &                         - \\
25 &      Mn &          194 &                   6.91 &       (2d) & - &                         - \\
26 &      Fe &          194 &                   6.73 &       (2c) & * &         exp.\cite{Mao1990-rd} \\
27 &      Co &          225 &                   6.67 &       (4a) & - &                         - \\
28 &      Ni &          225 &                   6.73 &       (4a) & - &                         - \\
29 &      Cu &          225 &                   6.91 &       (4a) & - &                         - \\
30 &      Zn &          194 &                   7.49 &       (2d) & - &                         - \\
31 &      Ga &          225 &                   8.27 &       (4a) & - &                         - \\
32 &      Ge &          194 &                   8.74 &       (2c) & * &     exp.\cite{Akahama2021-tw} \\
33 &      As &          229 &                   9.07 &       (2a) & - &                         - \\
34 &      Se &          139 &                   9.44 &       (2a) & * &     exp.\cite{Akahama2021-tw} \\
35 &      Br &          225 &                   9.91 &       (4a) & * &        th.\cite{Duan2010-yb} \\
36 &      Kr &          194 &                   11.09 &       (2d) & - &                         - \\
37 &      Rb &          194 &                   10.83 &       (2a) & lit. &          th.\cite{Ma2008-yy} \\
& & & & (2d)  & \\
38 &      Sr &          194 &                   10.61 &       (2c) & - &                         - \\
39 &       Y &  70 &  10.25 & (16e) x=-0.18749 & lit. &       th.\cite{Chen2012-xx} \\
40 &      Zr &          229 &                   10.05 &       (2a) & - &                         - \\
41 &      Nb &          229 &                   10.04 &       (2a) & * & th.\cite{Krasilnikov2014-pn} \\
42 &      Mo &          229 &                   9.99 &       (2a) & * & th.\cite{Krasilnikov2014-pn} \\
43 &      Tc &          194 &                   9.52 &       (2c) & * &        th.\cite{Shah2021-ul} \\
44 &      Ru &          194 &                   9.32 &       (2d) & * &         th.\cite{Liu2020-bg} \\
45 &      Rh &          225 &                   9.30 &       (4a) & - &                         - \\
46 &      Pd &          225 &                   9.43 &       (4a) & - &                         - \\
47 &      Ag &          225 &                   9.76 &       (4a) & - &                         - \\
48 &      Cd &          194 &                   10.52 &       (2c)& - &                         - \\
49 &      In &          139 &                   11.38 &       (2a) & * &       th.\cite{Simak2000-gx} \\
50 &      Sn &          194 &                   12.01 &       (2d) & - &                         - \\
51 &      Sb &          229 &                   12.37 &       (2a) & - &                         - \\
52 &      Te &          225 &                   12.49 &       (4a) & - &                         - \\
53 &       I &          139 &                   12.73 &       (2b) & - &                         - \\
54 &      Xe &          194 &                   14.19 &       (2c) & - &                         - \\
55 &      Cs &          225 &                   13.87 &       (4a) & - &                         - \\
56 &      Ba &          194 &                   14.03 &       (2c) & - &                         - \\
57 &      La &          139 &                   12.61 &       (2a) & - &                         - \\
\bottomrule
\end{tabular}
\caption{\label{tab:300gs}Ground-state database (DB\_GS) at 300 GPa. The symbols in the \textit{Source} column indicate: 
(i) \emph{*}  the structure is \emph{matching}; (ii) \emph{-}  no  reference could be found in literature; (iii) \emph{lit.}/\emph{ea.} the 
ground-state 
structure originates from literature/evolutionary algorithm. In the 
\emph{Ref} column, (i) \emph{th.} (\emph{exp.}) specifies whether the literature source is computational (experimental), or
(ii) \emph{-} missing.}
\end{table}



\begin{table}[ht]
\centering
\fontsize{6.5}{8}\selectfont
\begin{tabular}{l l l l >{\em}l l l}
\toprule
 Z & Element &  Space group &  Volume &  \normalfont{Wyckoff positions} & {$\Delta$H$_\text{EA-GS}$} &             Ref \\
 & & & (\AA$^3/atom$) & & (meV/atom) & \\
\midrule
 1 &       H &            4 &                   12.19 &      (2a) x=0.05164 y=0.04451 z=-0.20950 & <26 & exp.\cite{Vindryavskyi1980-cn} \\
 & & & & (2a) x=-0.44063 y=0.45391 z=-0.30440  & \\
  & & & & (2a) x=-0.04582 y=-0.05913 z=-0.28855  & \\
  & & & & (2a) x=0.46408 y=-0.45833 z=-0.20914  & \\
 2 &      He &          194 &                   17.30 &      (2c) & 0 &      exp.\cite{Henshaw1958-wj} \\
 3 &      Li &          139 &          20.20 &      (2b) &  <26 &       exp.\cite{Barrett1956-sw} \\
 4 &      Be &          194 &                   7.91 &      (2c) & 0 &       exp.\cite{Larsen1984-hg} \\
 5 &       \bf{B} &          166 &                   7.33 &      (6c) z=-0.06626 & 175.9 &      exp.\cite{Decker1959-qb} \\
 & & & & (6c) z=-0.19851  & \\
 6 &       C &          191 &                   10.34 &      (2d) & 0 &       exp.\cite{Zemann1965-mf} \\
 7 &       N &            4 &                   27.47 &     (2a) x=0.10164   y=-0.21303    z=0.37309 & <26 &      exp.\cite{Schuch1970-tj} \\
 & & & & (6c) x=-0.39456   y=-0.27416   z=-0.36853  & \\
 & & & & (6c) x=0.38647    y=0.28405    z=0.14917  & \\
 & & & & (2a) x=-0.10394    y=0.23297   z=-0.15331  & \\
 8 &       O &           13 &                   13.71 &      (4g) x=0.14641 y=0.17573 z=0.28949 & <26 &       exp.\cite{Datchi2014-pf} \\
 9 &       F &           15 &                  18.36 &      (8f) x=-0.39931 y=0.03411 z=0.00111 & <26 &        exp.\cite{Meyer1968-on} \\
 & & & & (8f) x=0.46870 y=0.39850 z=0.49744  & \\
10 &      Ne &          225 &                   22.36 &      (4a) & 0 &       exp.\cite{Zemann1965-mf} \\
11 &      Na &          225 &                   38.46 &      (4a) & <26 &       exp.\cite{Zemann1965-mf} \\
12 &      Mg &          194 &                   23.09 &      (2d) & 0 &       exp.\cite{Zemann1965-mf} \\
13 &      Al &          225 &                   16.53 &      (4a) & 0 &       exp.\cite{Zemann1965-mf} \\
14 &      Si &          227 &                   20.46 &      (8b) & 0 &       exp.\cite{Zemann1965-mf} \\
15 &       P &          164 &                   23.84 &      (2d) z=0.12144 & 0 &        exp.\cite{Cartz1979-wd} \\
16 &       \bf{S} &           82 &                   37.93 &      (8g) x=0.34946 y=0.21652 z=0.19856 & 32.4 &       exp.\cite{Rettig1987-em} \\
& & & & (8g) x=0.27671 y=0.35174 z=0.42027  & \\
17 &      Cl &           12 &                   42.58 &      (4i) x=-0.11504 z=-0.29223 & <26 &       exp.\cite{Powell1984-tf} \\
18 &      Ar &          225 &                   45.84 &      (4a) & 0 &       exp.\cite{Zemann1965-mf} \\
19 &       K &          194 &                   73.37 &      (2d) & <26 &       exp.\cite{Zemann1965-mf} \\
20 &      Ca &          225 &                   42.58 &      (4a) & 0 &       exp.\cite{Zemann1965-mf} \\
21 &      Sc &          194 &                   24.52 &      (2d) & 0 &       exp.\cite{Zemann1965-mf} \\
22 &      Ti &          194 &                   17.42 &      (2c) & 0 &       exp.\cite{Zemann1965-mf} \\
23 &       V &          229 &                   13.50 &      (2a) & 0 &       exp.\cite{Zemann1965-mf} \\
24 &      Cr &          229 &                   11.46 &      (2a) & 0 &       exp.\cite{Zemann1965-mf} \\
25 &      \bf{Mn} &          223 &                   10.69 &      (2a) & 62.7 &  exp.\cite{Oberteuffer1970-ll} \\
& & & & (6d)  & \\
26 &      Fe* &          229 &                   11.77 &      (2a) & 0 &         exp.\cite{Wilburn1978} \\
27 &      Co* &          194 &                   10.20 &      (2c) & 0 &       exp.\cite{Zemann1965-mf} \\
28 &      Ni* &          225 &                   10.76 &      (4a) & 0 &       exp.\cite{Zemann1965-mf} \\
29 &      Cu &          225 &                   12.00 &      (4a) & 0 &       exp.\cite{Zemann1965-mf} \\
30 &      Zn &          194 &                   15.25 &      (2d) & 0 &       exp.\cite{Zemann1965-mf} \\
31 &      Ga &           64 &                   20.53 &      (8f) y=0.34222 z=0.41786 & 0 &       exp.\cite{Sharma1962-sm} \\
32 &      Ge &          227 &                   24.15 &      (8a) & 0 &          exp.\cite{Hom1975-bu} \\
33 &      As &          166 &                   22.93 &      (6c) & 0 &     exp.\cite{Schiferl1969-ku} \\
34 &      Se &            5 &                   38.51 &      (4c) x=0.29680 y=0.21693 z=-0.09413 & <26 &       exp.\cite{Cherin1972-ej} \\
& & & & (4c) x=-0.34697 y=0.45560 z=0.24726  & \\
& & & & (4c) x=-0.10462 y=0.47555 z=-0.24530  & \\
& & & & (4c) x=-0.18926 y=0.46879 z=0.39753  & \\
35 &      Br\textsuperscript{\textdagger} &         62 & 36.16 & (4c) x=-0.25210 z=-0.09056 & <26 &  exp.\cite{Powell1984-tf} \\
& & & & (4c) x=-0.00218 z=0.24949  & \\
36 &      Kr &          225 &                   66.44 &      (4a) & 0 &       exp.\cite{Zemann1965-mf} \\
37 &      Rb &          139 &                   184.31 &      (2a) & <26 &       exp.\cite{Zemann1965-mf} \\
38 &      Sr &          225 &                   54.97 &      (4a) & 0 &       exp.\cite{Zemann1965-mf} \\
39 &       Y &          194 &                   32.53 &      (2c) & 0 &       exp.\cite{Zemann1965-mf} \\
40 &      Zr &          194 &                   23.44 &      (2d) & 0 &       exp.\cite{Zemann1965-mf} \\
41 &      Nb &          229 &                   18.16 &      (2a) & 0 &       exp.\cite{Zemann1965-mf} \\
42 &      Mo &          229 &                   15.83 &      (2a) & 0 &       exp.\cite{Zemann1965-mf} \\
43 &      Tc &          164 &                   14.47 &      (2d) z=-0.25317 & <26 &                          exp.\cite{Zemann1965-mf} \\
44 &      Ru &          194 &                   13.71 &      (2c) & 0 &       exp.\cite{Zemann1965-mf} \\
45 &      Rh &          225 &                   14.08 &      (4a) & 0 &       exp.\cite{Zemann1965-mf} \\
46 &      Pd &          225 &                   15.30 &      (4a) & 0 &       exp.\cite{Zemann1965-mf} \\
47 &      Ag &          225 &                   18.00 &      (4a) & 0 &          exp.\cite{Suh1988-rz} \\
48 &      Cd &          166 &                   23.10 &      (3a) & <26 &       exp.\cite{Zemann1965-mf} \\
49 &      In &          229 &                   27.74 &      (2a) & <26 &        exp.\cite{Smith1964-gm} \\
50 &      Sn &          227 &                   36.87 &      (8a) & 0 &        exp.\cite{Smith1964-gm} \\
51 &      Sb &          166 &                   32.13 &      (6c) z=0.26654 & 0 &     exp.\cite{Schiferl1969-ku} \\
52 &      \bf{Te} &          166 &                   32.56 &      (3a) & 46.5 &       exp.\cite{Adenis1989-jc} \\
53 &       I\textsuperscript{\textdagger} &           63 & 42.09 & (4a) & <26 &  exp.\cite{Zemann1965-mf} \\
& & & & (4c) y=-0.36584  & \\
54 &      Xe &          225 &                   88.43 &      (4a) & 0 &       exp.\cite{Zemann1965-mf} \\
55 &      Cs &          166 &                   113.82 &      (3a) & <26 &       exp.\cite{Zemann1965-mf} \\
56 &      Ba &          229 &                   64.07 &      (2a) & 0 &       exp.\cite{Zemann1965-mf} \\
57 &      La &          225 &                   37.85 &      (4a) & <26 &       exp.\cite{Zemann1965-mf} \\
\bottomrule
\end{tabular}
\caption{\label{tab:000ea}Evolutionary algorithm database (DB\_EA) at 0 GPa. In the \emph{$\Delta$H$_\text{EA-GS}$} column, (i) \emph{0}
indicates that  the EA crystal structure is lower in enthalphy than the corresponding LIT crystal structure; (ii) \emph{<26} that the EA crystal structure is degenerate in enthalpy with the corresponding LIT crystal structure,  (iii) Otherwise, it indicates the difference in enthalpy between the ground-state crystal structure and the EA-generated crystal structure, in meV/atom.  In the \emph{Ref} column, (i) \emph{th.} (\emph{exp.}) indicates that the literature source is  computational (experimental), or (ii) \emph{-} missing; (iii) \emph{Fe*}, \emph{Co*} and \emph{Ni*} indicate that the calculation is spin-polarized, with a magnetic moment of 2.22, 1.74, 0.606 $a.u.$ respectively\cite{Galperin1978-dz} and (iv) {$Br^\dag$} and {$I^\dag$}, that the calculation includes vdW interactions through the opt88-vdW exchange-correlation functional \cite{vdw_michaelides_2010}.}
\end{table}


\begin{table}[ht]
\centering
\fontsize{7}{8.5}\selectfont
\begin{tabular}{l l l l >{\em}l l l}
\toprule
 Z & Element &  Space group &  Volume &  \normalfont{Wyckoff positions} & {$\Delta$H$_\text{EA-GS}$} &             Ref \\
 & & & (\AA$^3/atom$) & & (meV/atom) & \\
\midrule
1 &       H &          62 &                   2.31 &     (4c) x=-0.24849 z=-0.10890 & <26 &                             th.\cite{Pickard2007-lq} \\
     & & & & (4c) x=-0.41675 z=0.14571  & \\
 2 &      He &          194 &                   3.46 &      (2c) & 0 &                                                 - \\
 3 &      Li &           43 &                   6.06 &      (16b) x=0.37357 y=0.47947 z=0.47349 & <26 &                                   th.\cite{Lv2011-gt} \\
    & & & & (16b) x=0.25385 y=0.12473 z=0.20675  & \\
 4 &      Be &          194 &                   5.27 &      (2c) & 0 &                                th.\cite{Kadas2007-ek} \\
 5 &       B &           64 &                   5.00 &      (8f) y=-0.15747 z=-0.08761 & 0 &                    exp.\cite{Shirai2011-xb} \\
 6 &       C &          227 &                   4.83 &      (8b) & 0 &                   th.\cite{Grumbach1996-ak} \\
 7 &       N &          199 &                   5.52 &      (8a) x=0.32248 & 0 &                                 th.\cite{Wang2010-ct} \\
 8 &       O &           15 &                   6.00 &      (8f) x=0.09795 y=0.29000 z=0.21503 & 0 &                                   th.\cite{Ma2007-vj} \\
 9 &       F &           64 &                   6.29 &      (8f) y=0.34595 z=0.11698 & 0 &                                th.\cite{Olson2020-ng} \\
10 &      Ne &          225 &                   6.71 &      (4a) & 0 &                              exp.\cite{Dewaele2008-qe} \\
11 &      Na &          225 &                   10.78 &      (4a) & 0 &                             exp.\cite{Hanfland2002-oc} \\
12 &      Mg &          229 &                   11.25 &      (2a) & 0 &                                   th.\cite{Li2010-hy} \\
13 &      Al &          225 &                   10.34 &      (4a) & 0 &                               exp.\cite{Fiquet2019-wc} \\
14 &      Si &          225 &                   9.19 &      (4a) &  0 &                              exp.\cite{Mujica2003-nr} \\
15 &       P &          221 &                   9.95 &      (1a) & 0 &                              exp.\cite{Akahama1999-yg} \\
16 &       \bf{S} &          166 &                   9.99 &      (3b) &  0 &                             exp.\cite{Whaley-Baldwin2020-vs} \\
17 &      Cl &           64 &                   11.40 &      (8f) y=0.31818 z=-0.37903 &  0 &                               th.\cite{Olson2020-ng} \\
18 &      Ar &          194 &                   12.62 &      (2d) & 0 &                                                  - \\
19 &       K &           64 &                   11.81 &      (8d) x=-0.28496 &  0 &                                  th.\cite{Ma2008-yy} \\
& & & & (8f) y=0.32494 z=0.32775  & \\
20 &      Ca &           92 &                   12.52 &      (8b) x=-0.18001 y=-0.48461 z=-0.15417 & <26 &             th.\cite{Oganov2010-my} \\
21 &      Sc &           88 &                   12.11 &      (16f) z=-0.27743 y=-0.19917 z=0.47208 & 0 &                              exp.\cite{Akahama2005-yi} \\
22 &      Ti &          139 &                   10.89 &      (2b) &  <26 &                             th.\cite{Kutepov2003-jg} \\
23 &       V &          229 &                   9.85 &      (2a) & 0 &                                th.\cite{Verma2008-on} \\
24 &      Cr &          229 &                   9.06 &      (2a) & 0 &                                                  - \\
25 &      Mn &          194 &                   8.48 &      (2c) & 0 &                          exp.\cite{Magad-Weiss2020-pd} \\
26 &      Fe &          194 &                   8.18 &      (2c) &  0 &                                 exp.\cite{Mao1990-rd} \\
27 &      Co &          225 &                   8.15 &      (4a) & <26 &                                  exp.\cite{Yoo2000-dz} \\
28 &      Ni &          225 &                   8.31 &      (4a) & 0 &                         th.\cite{Belashchenko2020-od} \\
29 &      Cu &          225 &                   8.67 &      (4a) & 0 &                                                  - \\
30 &      Zn &          194 &                   9.64 &      (2c) & 0 &                              exp.\cite{Akahama2021-tw} \\
31 &      Ga &          225 &                   10.93 &      (4a) & 0 &                              th.\cite{Simak2000-gx} \\
32 &      Ge &          139 &                   11.64 &      (2a) & <26 &                              exp.\cite{Akahama2021-tw} \\
33 &      As &          229 &                   12.02 &      (2a) & 0 &                                th.\cite{Silas2008-af} \\
34 &      Se &          166 &                   12.72 &      (3b) & 0 &                           exp.\cite{Degtyareva2005-bf} \\
35 &      Br &           71 &                   13.68 &      (2d) & <26 &                                   th.\cite{Li2020-pg} \\
36 &      Kr &          194 &                   15.58 &      (2c) & 0 &                                                  - \\
37 &      Rb &           64 &                   15.00 &      (8d) x=0.21706 & 0 &                           th.\cite{Ma2008-yy} \\
& & & & (8f) y=-0.32320 z=-0.17520  & \\
38 &      Sr &           62 &                   14.96 &      (4c) x=0.17310 z=-0.07136 & 0 &                                                  - \\
39 &       Y &           14 &                   14.80 &      (4e) x=0.24843 y=0.04533 z=-0.28604 & <26 &  exp.\cite{Pace2020-ih} \\
40 &      Zr &          229 &                   13.82 &      (2a) & 0 &                           exp.\cite{Anzellini2020-uc} \\
41 &      Nb &          229 &                   13.00 &      (2a) & 0 &                          th.\cite{Krasilnikov2014-pn} \\
42 &      Mo &          229 &                   12.48 &      (2a) & 0 &                          th.\cite{Krasilnikov2014-pn} \\
43 &      Tc &          194 &                   11.73 &      (2d) & 0 &                                 th.\cite{Shah2021-ul} \\
44 &      Ru &          194 &                   11.26 &      (2c) & 0 &                                  th.\cite{Liu2020-bg} \\
45 &      Rh &          225 &                   11.26 &      (4a) & 0 &                                                  - \\
46 &      Pd &          225 &                   11.61 &      (4a) & 0 &                                    th.\cite{National_University_of_Defense_Technology_Changsha2021-nk} \\
47 &      Ag &          225 &                   12.28 &      (4a) & 0 &                                                  - \\
48 &      Cd &          194 &                   13.52 &      (2d) & 0 &                                                  - \\
49 &      \bf{In} &          225 &                   15.07 &      (4a) & 0 &                                th.\cite{Simak2000-gx} \\
50 &      Sn &          139 &                   15.89 &      (2b) &  0 &                                  th.\cite{Yu2006-cz} \\
51 &      Sb &          229 &                   16.39 &      (2a) & 0 &                                                  - \\
52 &      Te &          225 &                   16.53 &      (4a) & 0 &                                  th.\cite{Liu2018-gy} \\
53 &       I &          225 &                   17.35 &      (4a) & 0 &                                                  - \\
54 &      Xe &          194 &                   20.05 &      (2c) & 0 &                                                  - \\
55 &      Cs &          194 &                   19.00 &      (2a) & 0 &                                 th.\cite{Guan2020-lt} \\
& & & & (2c)  & \\
56 &      Ba &          194 &                   18.73 &      (2d) & 0 &                                                  - \\
57 &      La &          225 &                   16.73 &      (4a) & 0 &                                 exp.\cite{Chen2022-wm} \\
\bottomrule
\end{tabular}
\caption{\label{tab:100ea}Evolutionary algorithm database (DB\_EA) at 100 GPa. In the \emph{$\Delta$H$_\text{EA-GS}$} column, (i) \emph{0}
indicates that  the EA crystal structure is lower in enthalphy than the corresponding LIT crystal structure; (ii) \emph{<26} that the EA crystal structure is degenerate in enthalpy with the corresponding LIT crystal structure,  (iii) Otherwise, it indicates the difference in enthalpy between the ground-state crystal structure and the EA-generated crystal structure, in meV/atom.  In the \emph{Ref} column, (i) \emph{th.} (\emph{exp.}) indicates that the literature source is  computational (experimental), or (ii) \emph{-} missing.}
\end{table}


\begin{table}[ht]
\centering
\fontsize{7}{8.5}\selectfont
\begin{tabular}{l l l l >{\em}l l l}
\toprule
 Z & Element &  Space group &  Volume &  \normalfont{Wyckoff positions} & {$\Delta$H$_\text{EA-GS}$} &             Ref \\
 & & & (\AA$^3/atom$) & & (meV/atom) & \\
\midrule
 1 &       H &           64 &                   1.70 &      (8f) y=-0.12910 z=0.03916 & <26 &      th.\cite{Pickard2007-lq} \\
 2 &      He &          194 &                   2.70 &      (2c) & 0 &                          - \\
 3 &      Li &          230 &                   4.52 &      (16b) &  <26 &           th.\cite{Lv2011-gt} \\
 4 &      Be &          194 &                   4.33 &      (2d) & 0 &        th.\cite{Kadas2007-ek} \\
 5 &       B &           64 &                   4.34 &      (8f) y=0.34220 z=0.41304 & 0 &       exp.\cite{Shirai2011-xb} \\
 6 &       C &          227 &                   4.34 &      (8b) & 0 &     th.\cite{Grumbach1996-ak} \\
 7 &       \bf{N} &          113 &                   4.64 &      (4e) x=-0.33587 z=0.32228 & 0 &         th.\cite{Wang2010-ct} \\
 & & & & (4d) z=-0.16015   & \\
 8 &       O &           12 &                   5.08 &      (4i) x=0.22601 z=-0.20344 & 0 &           th.\cite{Ma2007-vj} \\
 & & & & (4i) x=-0.29538 z=-0.20316   & \\
 & & & & (8j) x=-0.46534 y=0.25994 z=0.20327  & \\
 9 &       F &           64 &                   5.28 &      (8f) y=-0.16484 z=-0.11903 & 0 &        th.\cite{Olson2020-ng} \\
10 &      Ne &          225 &                   5.54 &      (4a) & 0 &      exp.\cite{Dewaele2008-qe} \\
11 &     Na &           62 &                   8.00 &      (4c) x=0.32391 z=-0.08020 & <26 &           exp.\cite{Ma2009-ey} \\
& & & & (4c) x=0.48576 z=0.30971   & \\
12 &      Mg &          229 &                   9.02 &      (2a) & 0 &           th.\cite{Li2010-hy} \\
13 &      Al &          194 &                   8.45 &      (2c) & <26 &       exp.\cite{Fiquet2019-wc} \\
14 &      Si &          225 &                   7.81 &      (4a) & 0 &       exp.\cite{Mujica2003-nr} \\
15 &       P &          191 &                   7.90 &      (1b) & 0 &      exp.\cite{Akahama2000-sg} \\
16 &       S &          166 &                   8.13 &      (3b) & 0 &          exp.\cite{Luo1993-pi} \\
17 &      Cl &           71 &                   9.07 &      (2a) & 0 &        th.\cite{Olson2020-ng} \\
18 &      Ar &          194 &                   10.20 &     (2c) & 0 &                          - \\
19 &       K &           64 &                   9.73 &      (8d) x=-0.21662 & 0 &           th.\cite{Ma2008-yy} \\
& & & & (8f) y=0.32384 z=0.17522  & \\
20 &      Ca &           62 &                   9.30 &      (4c) x=0.33247 z=-0.39914 & 0 &       exp.\cite{Sakata2011-qz} \\
21 &      \bf{Sc} &          14 &                   9.22 &     (4e) x=-0.24789 y=-0.08441 z=0.29798 & 0 &      exp.\cite{Akahama2005-yi} \\
22 &      Ti &          229 &                   8.72 &      (2a) & 0 &      th.\cite{Kutepov2003-jg} \\
23 &       V &          166 &                   8.39 &      (3a) & 0 &        th.\cite{Verma2008-on} \\
24 &      Cr &          229 &                   7.97 &      (2a) & 0 &                          - \\
25 &      Mn &          194 &                   7.57 &      (2d) & 0 &  exp.\cite{Magad-Weiss2020-pd} \\
26 &      Fe &          194 &                   7.29 &      (2c) & 0 &          exp.\cite{Mao1990-rd} \\
27 &      Co &          225 &                   7.24 &      (4a) & 0 &          exp.\cite{Yoo2000-dz} \\
28 &      \bf{Ni} &          225 &                   7.34 &      (4a) & 0\textsuperscript{\textdagger} & th.\cite{Belashchenko2020-od} \\
29 &      Cu &          225 &                   7.57 &      (4a) & 0 &                          - \\
30 &      Zn &          194 &                   8.28 &      (2c) & 0 &      exp.\cite{Akahama2021-tw} \\
31 &      Ga &          225 &                   9.21 &      (4a) & 0 &        th.\cite{Simak2000-gx} \\
32 &      Ge &          194 &                   9.80 &      (2c) & 0 &      exp.\cite{Akahama2021-tw} \\
33 &      As &          229 &                   10.15 &      (2a) & 0 &        th.\cite{Silas2008-af} \\
34 &      Se &          229 &                   10.62 &      (2a) & 0 &      exp.\cite{Akahama2021-tw} \\
35 &      Br &          225 &                   11.20 &      (4a) & 0 &           th.\cite{Li2020-pg} \\
36 &      Kr &          194 &                   12.63 &      (2d) & 0 &                          - \\
37 &      Rb &          194 &                   12.18 &      (2a) & 0 &           th.\cite{Ma2008-yy} \\
& & & & (2d)  & \\
38 &      Sr &          194 &                   11.92 &      (2d) & 0 &                          - \\
39 &       Y &           70 &                   11.64 &      (16e) x=0.43748 & 0 &       th.\cite{LI2019-x} \\
40 &      Zr &          229 &                   11.32 &      (2a) & 0 &                          - \\
41 &      Nb &          229 &                   11.17 &      (2a) & 0 &  th.\cite{Krasilnikov2014-pn} \\
42 &      Mo &          229 &                   10.97 &      (2a) & 0 &  th.\cite{Krasilnikov2014-pn} \\
43 &      Tc &          194 &                   10.35 &      (2c) & 0 &                          - \\
44 &      Ru &          194 &                   10.09 &      (2c) & 0 &          th.\cite{Liu2020-bg} \\
45 &      Rh &          225 &                   10.06 &      (4a) & 0 &                          - \\
46 &      Pd &          225 &                   10.25 &      (4a) & 0 &                          - \\
47 &      Ag &          225 &                   10.69 &      (4a) & 0 &                          - \\
48 &      Cd &          194 &                   11.50 &      (2c) & 0 &                          - \\
49 &      In &          139 &                   12.68 &      (2b) & 0 &        th.\cite{Simak2000-gx} \\
50 &      Sn &          194 &                   13.40 &      (2c) & 0 &           th.\cite{Yu2006-cz} \\
51 &      Sb &          229 &                   13.80 &      (2a) & 0 &                          - \\
52 &      Te &          225 &                   13.96 &      (4a) & 0 &          th.\cite{Liu2018-gy} \\
53 &       I &          225 &                   14.37 &      (4a) & 0 &                          - \\
54 &      Xe &          194 &                   16.24 &      (2c) & 0 &                          - \\
55 &      Cs &          225 &                   15.60 &      (4a) & 0 &         th.\cite{Guan2020-lt} \\
56 &      Ba &          194 &                   15.67 &      (2d)& 0 &                          - \\
57 &      La &          225 &                   14.14 &      (4a) & 0 &        exp.\cite{Chen2022-wm} \\
\bottomrule
\end{tabular}
\caption{\label{tab:200ea}Evolutionary algorithm database (DB\_EA) at 200 GPa. In the \emph{$\Delta$H$_\text{EA-GS}$} column, (i) \emph{0}
indicates that  the EA crystal structure is lower in enthalphy than the corresponding LIT crystal structure; (ii) \emph{<26} that the EA crystal structure is degenerate in enthalpy with the corresponding LIT crystal structure,  (iii) Otherwise, it indicates the difference in enthalpy between the ground-state crystal structure and the EA-generated crystal structure, in meV/atom.  In the \emph{Ref} column, (i) \emph{th.} (\emph{exp.}) indicates that the literature source is  computational (experimental), or (ii) \emph{-} missing.}
\end{table}


\begin{table}[ht]
\centering
\fontsize{7}{8.5}\selectfont
\begin{tabular}{l l l l >{\em}l l l}
\toprule
 Z & Element &  Space group &  Volume &  \normalfont{Wyckoff positions} & {$\Delta$H$_\text{EA-GS}$} &             Ref \\
 & & & (\AA$^3/atom$) & & (meV/atom) & \\
\midrule
 1 &       H &           64 &                   1.44 &       (8f) y=-0.12912 z=-0.05825 & <26 &     th.\cite{Pickard2007-lq} \\
 2 &      He &          194 &                   2.32 &       (2d) & 0 &                         - \\
 3 &      \bf{Li} &           73 &                   3.82 &       (16f) x=0.12476 y=0.12417 z=0.37634 & 40.3 &          th.\cite{Lv2011-gt} \\
 4 &      Be &          194 &                   3.81 &       (2c) & 0 &          th.\cite{Lu2015-aa} \\
 5 &       B &           64 &                   3.93 &       (8f) y=0.157750 z=-0.08527 & 0 &                         - \\
 6 &       C &          194 &                   4.00 &       (4e) z=-0.09278 & 0 &                         - \\
  & & & & (4f) z=-0.34395  & \\
 7 &       \bf{N} &           64 &                   4.18 &       (8f) y= 0.10538 z=-0.10004 & 0 &        th.\cite{Wang2010-ct} \\
 8 &       O &          166 &                   4.55 &       (6c) z=-0.06915 & 0 &                         - \\
 9 &       F &           64 &                   4.72 &       (8f) y=-0.32735 z=-0.37974 & 0 &       th.\cite{Olson2020-mk} \\
10 &      Ne &          225 &                   4.91 &       (4a) & 0 &                         - \\
11 &      Na &          194 &                   6.68 &       (2a) & <26 &        exp.\cite{Ma2009-ey} \\
& & & & (2d)  & \\
12 &      Mg &          229 &                   7.84 &       (2a) & 0 &          th.\cite{Li2010-hy} \\
13 &      Al &          194 &                   7.57 &       (2c) & 0 &     exp.\cite{Akahama2006-gw} \\
14 &      Si &          225 &                   7.00 &       (4a) & 0 &                         - \\
15 &       P &          220 &                   6.84 &       (16c) x=0.47267 & 0 &                         - \\
16 &       S &          166 &                   7.15 &       (3a) & 0 &  th.\cite{Degtyareva2007-mq} \\
17 &      Cl &          139 &                   7.90 &       (2a) & 0 &       th.\cite{Olson2020-ng} \\
18 &      Ar &          194 &                   8.92 &       (2c) & 0 &                         - \\
19 &       K &          194 &                   8.47 &       (2a) & 0 &          th.\cite{Ma2008-yy} \\
& & & & (2d)  & \\
20 &      Ca &           62 &                   8.07 &       (4c) & 0 &      - \\
21 &      Sc &           70 &                   7.81 &       (16g) z=0.31244 & <26 &     exp.\cite{Akahama2005-yi} \\
22 &      Ti &          229 &                   7.61 &       (2a) & 0 &       exp.\cite{Zhang2022-jj} \\
23 &       V &          229 &                   7.57 &       (2a) & 0 &       th.\cite{Verma2008-on} \\
24 &      Cr &          229 &                   7.28 &       (2a) & 0 &                         - \\
25 &      Mn &          194 &                   6.91 &       (2d) & 0 &                         - \\
26 &      Fe &          194 &                   6.73 &       (2c) & 0 &         exp.\cite{Mao1990-rd} \\
27 &      Co &          225 &                   6.67 &       (4a) & 0 &                         - \\
28 &      Ni &          225 &                   6.73 &       (4a) & 0 &                         - \\
29 &      Cu &          225 &                   6.91 &       (4a) & 0 &                         - \\
30 &      Zn &          194 &                   7.49 &       (2d) & 0 &                         - \\
31 &      Ga &          225 &                   8.27 &       (4a) & 0 &                         - \\
32 &      Ge &          194 &                   8.74 &       (2c) & 0 &     exp.\cite{Akahama2021-tw} \\
33 &      As &          229 &                   9.07 &       (2a) & 0 &                         - \\
34 &      Se &          139 &                   9.44 &       (2a) & 0 &     exp.\cite{Akahama2021-tw} \\
35 &      Br &          225 &                   9.91 &       (4a) &  0 &       th.\cite{Duan2010-yb} \\
36 &      Kr &          194 &                   11.09 &       (2d) & 0 &                         - \\
37 &      Rb &          225 &              10.79 &    (4a) & <26 &          th.\cite{Ma2008-yy} \\
38 &      Sr &          194 &                   10.61 &       (2c) & 0 &                         - \\
39 &       \bf{Y} &           71 &                   10.24 &      (2c) & 110.1 &        th.\cite{Chen2012-xx} \\
40 &      Zr &          229 &                   10.05 &       (2a) & 0 &                         - \\
41 &      Nb &          229 &                   10.04 &       (2a) & 0 & th.\cite{Krasilnikov2014-pn} \\
42 &      Mo &          229 &                   9.99 &       (2a) & 0 & th.\cite{Krasilnikov2014-pn} \\
43 &      Tc &          194 &                   9.52 &       (2c) & 0 &        th.\cite{Shah2021-ul} \\
44 &      Ru &          194 &                   9.32 &       (2d) & 0 &         th.\cite{Liu2020-bg} \\
45 &      Rh &          225 &                   9.30 &       (4a) & 0 &                         - \\
46 &      Pd &          225 &                   9.43 &       (4a) &  0 &                        - \\
47 &      Ag &          225 &                   9.76 &       (4a) & 0 &                         - \\
48 &      Cd &          194 &                   10.52 &       (2c)& 0 &                         - \\
49 &      In &          139 &                   11.38 &       (2a) & 0 &       th.\cite{Simak2000-gx} \\
50 &      Sn &          194 &                   12.01 &       (2d) & 0 &                         - \\
51 &      Sb &          229 &                   12.37 &       (2a) & 0 &                         - \\
52 &      Te &          225 &                   12.49 &       (4a) & 0 &                         - \\
53 &       I &          139 &                   12.73 &       (2b) & 0 &                         - \\
54 &      Xe &          194 &                   14.19 &       (2c) & 0 &                         - \\
55 &      Cs &          225 &                   13.87 &       (4a) & 0 &                         - \\
56 &      Ba &          194 &                   14.03 &       (2c) & 0 &                         - \\
57 &      La &          139 &                   12.61 &       (2a) & 0 &                         - \\
\bottomrule
\end{tabular}
\caption{\label{tab:300ea}Evolutionary algorithm database DB\_EA at 300 GPa. In the \emph{$\Delta$H$_\text{EA-GS}$} column, (i) \emph{0}
indicates that  the EA crystal structure is lower in enthalphy than the corresponding LIT crystal structure; (ii) \emph{<26} that the EA crystal structure is degenerate in enthalpy with the corresponding LIT crystal structure,  (iii) Otherwise, it indicates the difference in enthalpy between the ground-state crystal structure and the EA-generated crystal structure, in meV/atom.  In the \emph{Ref} column, (i) \emph{th.} (\emph{exp.}) indicates that the literature source is  computational (experimental), or (ii) \emph{-}  missing.}
\end{table}


\begin{table}[ht]
\centering
\fontsize{7}{8.5}\selectfont
\begin{tabular}{l l l l l >{\em}l l l}
\toprule
Pressure & Z & Element &  Space group & Volume &  \normalfont{Wyckoff positions} & {$\Delta$H$_\text{LIT-GS}$} &                     Ref \\
\multicolumn{1}{c}{(GPa)} & & & & (\AA$^3/atom$) & & \normalfont{(meV/atom)} & \\
\midrule
\multicolumn{1}{c}{100} & 16 &       S &           bco & \emph{N/A} & \emph{N/A} & \emph{N/A} &  exp.\cite{Whaley-Baldwin2020-vs} \\
\multicolumn{1}{c}{100} & 49 &      In &           bct & \emph{N/A} & \emph{N/A} & \emph{N/A} &  th.\cite{Simak2000-gx} \\
\hline 
\multicolumn{1}{c}{200} & 7 & N & 32 & \emph{N/A} & \emph{N/A} & \emph{N/A} &   th.\cite{Wang2010-ct} \\
\multicolumn{1}{c}{200} & 21 & Sc & \emph{N/A} & \emph{N/A}  & \emph{N/A} & \emph{N/A} & exp.\cite{Akahama2005-yi} \\
\multicolumn{1}{c}{200} & 28 & Ni & 229 & 7.34 & (2a) & 173.2 & th.\cite{Belashchenko2020-od} \\
\hline 
\multicolumn{1}{c}{300} & 7 & N &  32 & \emph{N/A} & N/A & \emph{N/A} &              th.\cite{Wang2010-ct} \\
\bottomrule
\end{tabular}
\caption{\label{tab:miss}Database of structures (DB\_MISS) at different pressures.  The \emph{$\Delta$H$_\text{LIT-GS}$} column represents the difference in enthalpy between the ground-state crystal structure and the LIT crystal structure, in meV/atom.  In the \emph{Ref} column, (i) \emph{th.} (\emph{exp.}) specifies whether the literature source is computational (experimental).  N/A
indicates that the available information is too incomplete to completely characterize the structure ({\em unreproducible}).}
\end{table}

\end{document}